\documentclass[default,iicol]{sn-jnl}

\jyear{2021}%

\theoremstyle{thmstyleone}%
\theoremstyle{thmstyletwo}%

\theoremstyle{thmstylethree}%

\usepackage{epsfig,graphics,color,graphicx,amsmath}

\usepackage{inputenc}
\usepackage{soul}
\colorlet{darkred}{red!80!black}
\colorlet{darkgreen}{green!50!black}
\colorlet{darkblue}{blue!50!black}

\newcommand{\cd}{\makebox[0.08cm]{$\cdot$}}

\begin{document}

\title[\hspace{8.5cm}On cancellation of non-adiabatic and off-shell effects]{On cancellation of non-adiabatic and off-shell effects in the antiproton annihilation  in deuteron}

\author[1]{\fnm{O.D.} \sur{Dalkarov}}\email{odalkarov@rambler.ru}
\equalcont{These authors contributed equally to this work.}

\author[2]{\fnm{V.A.} \sur{Karmanov}}\email{karmanovva@lebedev.ru}
\equalcont{These authors contributed equally to this work.}

\author*[2]{\fnm{E.A.} \sur{Kupriyanova}}\email{kupr\_i\_k@mail.ru}
\equalcont{These authors contributed equally to this work.}

\affil[1]{\orgdiv{MEPhI}, \orgname{National Research Nuclear University}, \orgaddress{\street{Kashirskoe Highway 31}, \city{Moscow}, \postcode{115409}, 
\country{Russia}}}

\affil[2]{\orgdiv{Nuclear Physics and Astrophysics Division}, \orgname{Lebedev Physical Institute}, \orgaddress{\street{Leninsky Prospect 53}, \city{Moscow}, \postcode{119991}, 
\country{Russia}}}


\abstract{As known, some approximate approaches  to the hadron scattering from nuclei work rather well  far beyond the limits of their applicability. This was explained by cancellation of the contributions (non-adiabatic and off-shell effects) omitted in these approaches. Moreover, in some cases (in particular, for the reaction $\bar{p}d \to e^+e^-n$)  this cancellation allowed to derive rather simple analytical formula for the reaction amplitude.
Solving the Faddeev equations, we confirm  numerically this formula and, hence, the cancellations.}




\maketitle

\section{Introduction}\label{sec1}

A few decades ago, when numerical solving the Faddeev equations was not yet possible and calculation of the hadron scattering on nuclei based on approximate analytical approaches, it was noticed that some of these approximate approaches worked surprisingly well in wider domain than initially expected. As an example, we mention the Glauber approach \cite{Glauber}, 
which works well, often beyond the formal limits of its applicability. An illustration is given by good description of scattering of the low-energy antiprptons from nuclei, down to 50 MeV \cite{dk} (see for review Ref. \cite{dk87} and references therein). For applicability of this approach, two principal conditions should be satisfied: ({\it i}) straightforwardness of the trajectory of the projectile particle  in nucleus (eikonal approximation); ({\it ii}) possibility to neglect the motion of the intra-nuclear nucleons (adiabatic approximation). In addition, though the nucleons in the intermediate states are virtual (off-mass-shell),
in these calculations, ({\it iii}) the elementary nucleon-nucleon amplitudes were taken on-mass shell.  In the case of the low-energy antiproton scattering, the condition ({\it i})  can be ensured by the fact that the antiproton-nucleon scattering is strongly forward directed even for relatively low energies. However, for the first glance, the conditions  ({\it ii}) and ({\it iii}) can be hardly satisfied. 

At that time, these observations triggered the researches aimed to explain the reasons of these successes. The detail analytical study of the
hadron-nucleus 
 scattering amplitudes was fulfilled  \cite{KolKondr,KolKsen,DK81,DKK83,Faldt,Wallace,Gurvitz}.  The cancellation of different effects was discovered. Thus, it was found that the non-adiabatic effects and off-shell effects in the elementary scattering amplitudes considerably canceled each other. This explains, why, for example, the Glauber approach \cite{Glauber}, which just does not take into account these effects,  works so well.
 It also explains, why an approximate approach loses its accuracy after attempt to "improve" it by incorporating only a part of omitted effects, say, the non-adiabatic effects only, omitted in the eikonal propagators.

Currently, the problem of scattering of nucleons by light nuclei is successfully being solved by the numerical treatment of the Faddeev-Yakubovsky equations (see, e.g., the review \cite{RJ} for the equations and references therein for their solutions). The input is the nucleon-nucleon potential (or the off-shell amplitudes), the output is the many-channel scattering amplitude, which includes both the elastic scattering, rearrangements  and breakup, if any. 
This great and well-deserved success achieved, partially, due to the
"brute force" of computers,  overshadowed the interest in the important question "how does this work?"\, Though all the effects, discussed above, are automatically taken into account by the Faddeev-Yakubovsky equations, the question "how does this work?"\,, that is, "what is the role and contribution of the non-adiabatic and off-shell effects in forming the final many-channel scattering amplitude?"\, - still can be asked if one wants to achieve the deep understanding of mechanism of a nuclear reaction. 

\begin{figure}[h!]
\begin{center}
\includegraphics[width=7.5cm]{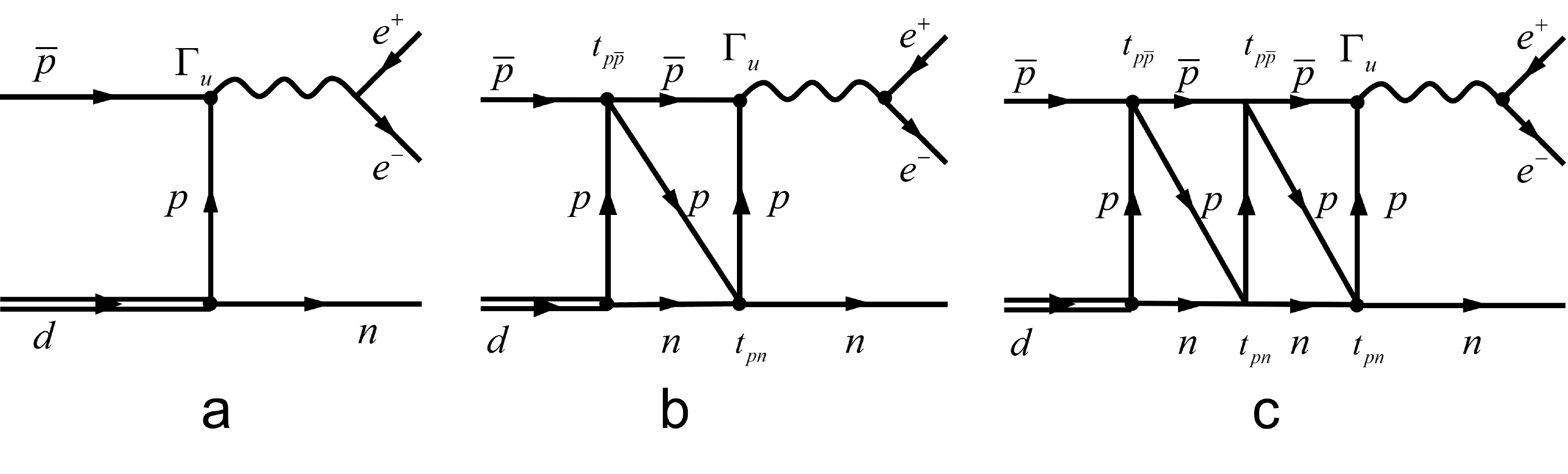}
\end{center}
\caption{Multiple scattering graphs determining amplitude of the reaction $\bar{p}d\to e^+e^-n$. }  \label{fig1}
\end{figure}

\begin{figure}[h!]
\begin{center}
\includegraphics[width=7cm]{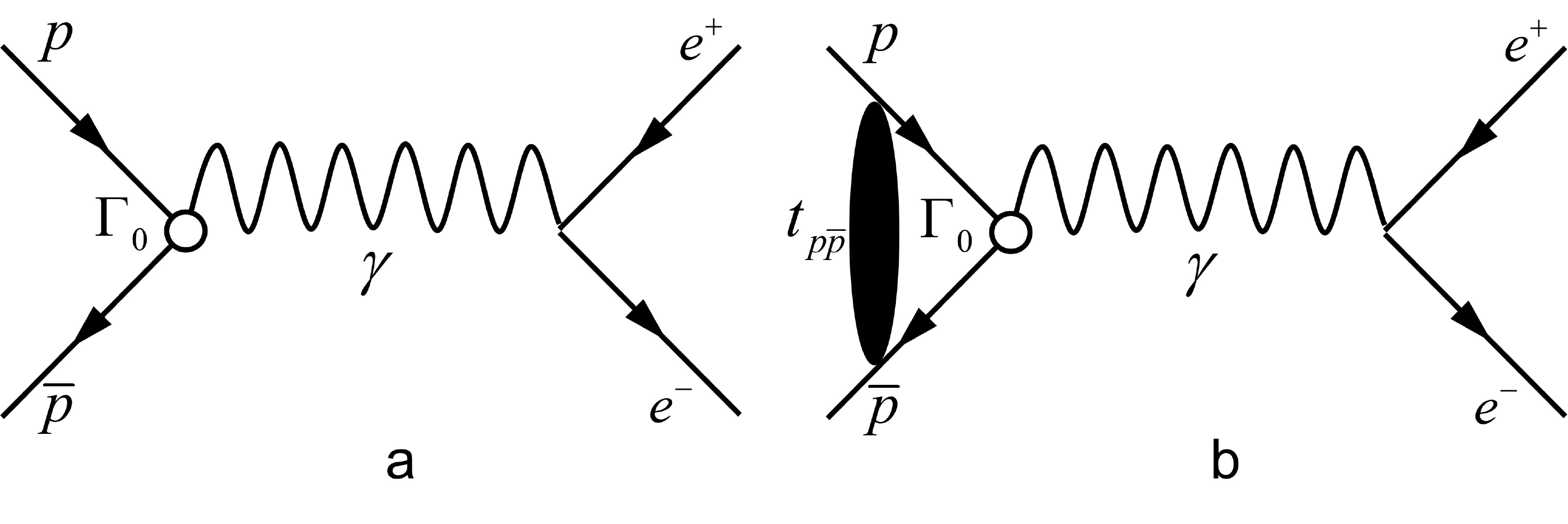}
\end{center}
\caption{The annihilation amplitude $\bar{p}p\to e^+e^-$.}  \label{fig2}
\end{figure}

These researches had not only academic, but also crucial practical interest, related to possibility to extract from
the reaction  $\bar{p}d\to e^+e^-n$ the proton form factor in the time-like domain 
of the momentum transfer squared $q^2>0$, under the $\bar{p}p$ threshold $q^2<4m^2$.
In the works  \cite{DK81,DKK83},  the scrutinized study of full set of the diagrams, shown in Fig. \ref{fig1}, contributing to the amplitude 
$\bar{p}d\to e^+e^-n$,  was carried out. The amplitude corresponding to the first term of this set, Fig. \ref{fig1}a, has the form
\begin{equation}\label{Mu}
M_u=\psi_d\Gamma_u,
\end{equation}
where $\psi_d$ is the deuteron wave function, $\Gamma_u$ is the annihilation amplitude $\bar{p}p\to e^+e^-$, Figure \ref{fig2},
\begin{equation}\label{Gamma_u}
\Gamma_u=\psi_{k_u}(r=0)\Gamma_0, 
\end{equation}
$\Gamma_0$ is the amplitude of annihilation $\bar{p}p\to e^+e^-$ not including the initial state $\bar{p}p$ interaction (Fig. \ref{fig2}a), and 
$\psi_{k}(r)$ is the continuous spectrum $\bar{p}p$ coordinate space wave function, corresponding to the eigenvalue $E=k^2/m$ ($k$ is the nucleon momentum in the $\bar{p}p$ c.m. frame, $m$ is the nucleon mass). It  takes into account  the initial state $\bar{p}p$ interaction, Fig. \ref{fig2}b. The equation (\ref{Gamma_u}) corresponds to  the sum of two graphs shown in Fig. \ref{fig2}. 

The subscript "$u$" (we follow the notations  of Refs.  \cite{DK81,DKK83})  denotes the $\bar{p}p$ energy found from the energy-momentum conservation laws applied to the reaction 
$\bar{p}d\to e^+e^-n$. It is calculated in the next section, Eq. (\ref{ucm}).
This energy, relative to the $\bar{p}p$ mass $2m$, can be both positive and negative.  That is, if the impulse approximation diagram Fig. \ref{fig1}a dominates, this indeed would allow to access the proton form factor under threshold. 

In contrast to Fig. \ref{fig2}b, which, besides the annihilation,  includes the initial state $\bar{p}p$ interaction, the antiproton-nucleon  interactions in the reaction $\bar{p}d \to e^+e^-n$, before the $\bar{p}p$ annihilation, are determined by the infinite set of the multiple scattering graphs shown in Fig. \ref{fig1}. These graphs are not reduced to a single $\bar{p}p$ interaction.
That's why, generally speaking, the impulse approximation formula (\ref{Mu}), proportional to the proton form factor, is distorted.   
However, as it was proved in \cite{DK81,DKK83},
the total effect of all the re-scatterings and cancellations mentioned above and shown in Fig. \ref{fig1}, restores the amplitude in the form of Eq. (\ref{Mu}), 
but with the shifted and always positive energy $E$ in $\psi_{k}(0)$, called below "$v$" (defined by Eq. (\ref{vcm})).  Namely,
\begin{equation}\label{Mv}
M_v=\psi_d\Gamma_v
\end{equation}
(which will be given in more detail in the next section) with
\begin{equation}\label{Gamma_v}
\Gamma_v=\psi_{k_v}(r=0)\Gamma_0.
\end{equation}

This approximate formula is not valid for $\bar{p}p$ in the vicinity of the $\bar{p}p$ bound states. Therefore, our discussion concerns the situation when the $e^+e^-$ effective mass is not close to mass of the $\bar{p}p$ bound states. The accuracy of Eq. (\ref{Gamma_v}) was estimated in \cite{DK81,DKK83} in 30\%.

Like in Eq. (\ref{Gamma_u}),  $\Gamma_0$ in Eq. (\ref{Gamma_v}) is still the amplitude of annihilation $\bar{p}p\to e^+e^-$ not including the initial state 
$\bar{p}p$ interaction. It contains the "internal" proton form factor which, like  in Eq. (\ref{Gamma_u}), depends on the true $q^2$, defined in the next section  by Eq. (\ref{qu2}). $q^2$ can take the under-threshold values less than $4m^2$. 
The argument $q^2$ is not shifted. However, though this formula looks like  the impulse approximation (Figure \ref{fig1}a),
the factor $\Gamma_v$, responsible for the initial state $\bar{p}p$ interaction, is quite different from  
$\Gamma_u$ in the impulse approximation, Eq. (\ref{Mu}).
It takes approximately into account an infinite set of diagrams corresponding to all the $np$ and $\bar{p}p$ re-scatterings shown in Figure \ref{fig1}. 
The energy $E$, denoted now, following Refs. \cite{DK81,DKK83}, by $v$, should be found not from the conservation laws, 
allowing to take the negative values. It takes the value as if the proton is not bound in the deuteron, but is free \cite{DK81,DKK83}.  That is, as if its four-momentum were on the nucleon mass shell. $v$ is always positive and therefore the $np$ and $\bar{p}p$ re-scatterings, resulting in Eq. (\ref{Gamma_v}), prevent from penetration under the $\bar{p}p$ threshold and don't allow to access the proton form factor in the time-like domain, except for the close vicinity of the $\bar{p}p$ bound states, if any.

This was very important and unexpected observation. For the first glance, the "common sense" intuition refuses to accept it. 
Therefore, it needs independent confirmation.
 The aim of the present work is to check this result not analyzing the diagrams (in contrast to the way taken in Refs. \cite{DK81,DKK83}),  but by direct numerical solving the Faddeev equations. In this our way, the results of the papers \cite{DK81,DKK83} will be confirmed with rather good precision (better than 30\% expected in \cite{DK81,DKK83}). 
 Similarly to these papers, we will consider the relatively small antiproton energies.  
 In the framework of the impulse approximation, the value of the cross section of the reaction $\bar{p}d \to e^+e^-n$ for high-energy antiprotons with $p=1.5$ GeV/c was estimated in \cite{Helene}.

Plan of the present article is the following. In Sect. \ref{sect2} we give the analytical formula for the amplitude $\bar{p}d \to e^+e^-n$ found in Refs.  \cite{DK81,DKK83}. In Sect. \ref{Fad} we present the formalism of the Faddeev equations in the coordinate space.
In Sect. \ref{ampl}  the amplitude $\bar{p}d \to e^+e^-n$ is calculated by means of the Faddeev formalism. The comparison of the
numerical results, found via the Faddeev equations, with Eqs. (\ref{Mu}) and (\ref{Mv}) is presented in Sect. \ref{num}. Sect. \ref{concl} contains concluding remarks. Some technical details are included in four Appendices \ref{Hulthen},\ref{Exp},\ref{Jacobi},\ref{unitarity}.

\section{Refining the amplitudes $M_u$ and $M_v$}\label{sect2}
The set of graphs determining the amplitude of the reaction $\bar{p}d \to e^+e^-n$ includes, before annihilation $\bar{p}p\to e^+e^-$, the re-scatterings of the incident antiproton $\bar{p}$ on nucleons of deuteron and the re-scatterings of the deuteron nucleons ($np$) on each other. Examples of these graphs taking into account the $\bar{p}p$ and $pn$ re-scatterings only (not the $\bar{p}n$ ones) are shown in Figure \ref{fig1}. The approximation omitting all these re-scatterings is the impulse approximation.  It takes into account only the graph Fig. \ref{fig1}a. We also omit, for two reasons, the Coulomb interaction between $\bar{p}$ and $p$. ({\it i})~Though we consider the low-energy antiprotons, their energy is not so low for importance of the Coulomb interaction. 
({\it ii}) Testing numerically the results of Refs. \cite{DK81,DKK83}, we have to assume exactly the same interaction which was taken in these papers. Otherwise, we will be unable to obtain clear conclusions.

The impulse approximation amplitude reads:
\begin{equation}\label{an3d}
M(\bar{p}d\to e^+e^- n)=\psi_d\left(\vec{p}_n+\frac{1}{2}\vec{p}_{\bar{p}}\right)\psi_{k_u}(0)\Gamma_0,
\end{equation}
where $k_u=\sqrt{m\,u}$ and $u$ is the $\bar{p}p$ energy (over or below $2m$) in this reaction, found from the conservation laws.
Its explicit form is given below in Eq. (\ref{ucm}). 
We work in the c.m. frame $\vec{p}_d+\vec{p}_{\bar{p}}=0$, where the amplitude depends on the momentum $\vec{p}_{\bar{p}}$ of the incident antiproton and on the momentum $\vec{p}_n$ of outgoing neutron and we assume that the proton and neutron masses are equal to each other: $m_p=m_n\equiv m$.
In Eq. (\ref{an3d}), $\vec{p}_n+\frac{1}{2}\vec{p}_{\bar{p}}$ is the relative $pn$ momentum in c.m. frame. Since 
the deuteron momentum is $\vec{p}_d=-\vec{p}_{\bar{p}}$, the proton momentum is $\vec{p}_p=\vec{p}_d-\vec{p}_n=-\vec{p}_{\bar{p}}-\vec{p}_n$, this gives  for the relative $np$ momentum $\frac{1}{2}(\vec{p}_n-\vec{p}_p)=\vec{p}_n+\frac{1}{2}\vec{p}_{\bar{p}}$.

The invariant $e^+e^-$ energy squared  (equal, for the graph Fig. \ref{fig1}a, to  the $\bar{p}p$ energy  squared)  is
\begin{eqnarray}
M^2_{e^+e^-}&=&q^2=(E_d+E_{\bar{p}}-E_n)^2-\vec{p}\,^2_n 
\label{qu2}\\
&=&\Bigl(\sqrt{(2m-\vert\epsilon_d\vert)^2+\vec{p}\,^2_{\bar{p}}}
+\sqrt{m^2+\vec{p}\,^2_{\bar{p}}}
\nonumber\\
&-&\sqrt{m^2+\vec{p}\,^2_n}\Bigr)^2-\vec{p}\,^2_n
\approx (2m+u)^2,
\nonumber
\end{eqnarray}
where $\epsilon_d$ is the deuteron binding energy and (in the non-relativistic limit -- large $m$):
\begin{equation}\label{ucm}
u=\frac{3(\vec{p}\,^2_{\bar{p}}-\vec{p}\,^2_n)}{4m}-\vert\epsilon_d\vert.
\end{equation}
This variable can take the negative under-threshold values.

As mentioned in the Introduction, according to   \cite{DK81,DKK83}, the amplitude corresponding to infinite sum of graphs shown in Fig. \ref{fig1},  due to cancellation in all of them the non-adiabatic effects in the nucleon propagators with the off-mass-shell effects in the elementary amplitudes,  obtains the form  of the impulse approximation  Eq. (\ref{an3d}), 
but with the shifted value of $k_u\to k_v= \sqrt{m\,v}$, where $v$  is the kinetic energy of the 
$\bar{p}p$ pair calculated as if the proton were on mass shell. Its explicit expression is given below by Eq. (\ref{vcm}).

We emphasize again: the proton in all the graphs Fig. \ref{fig1} is, of course, off-mass-shell.  
The shift of the argument $u\to v$, resulting to the prescription $k_v=\sqrt{m\,v}$, is due to incorporating the infinite sum of the graphs Figure \ref{fig1} and their partial cancellation, as it was proved in \cite{DK81,DKK83}.

Let us express now the energy $v$  via the momenta $\vec{p}_{\bar{p}}$  and $\vec{p}_n$.  The assumption that the proton is  on the mass shell means that we take for its energy not the difference $E_p=E_d-E_n$, but the value 
$E^{on}_p=\sqrt{m^2+ \vec{p}\,^2_p}$.
As stated above,  the proton momentum:
$\vec{p}_p=-\vec{p}_{\bar{p}}-\vec{p}_n$. 
The on-shell proton energy $E^{on}_p=\sqrt{m^2+ \vec{p}\,^2_p}=\sqrt{m^2+(\vec{p}_{\bar{p}}+\vec{p}_n)^2}$.
Then the invariant $\bar{p}p$ energy squared  is
\begin{eqnarray}
M^2_{e^+e^-}&=&(E^{on}_p+E_{\bar{p}})^2-(\vec{p}_p+\vec{p}_{\bar{p}})^2
\label{qv2}\\
&=&\left(\sqrt{m^2+(\vec{p}_{\bar{p}}+\vec{p}_n)^2}+\sqrt{m^2+\vec{p}\,^2_p}\right)^2
\nonumber\\
&-&\vec{p}\,^2_n \approx (2m+v)^2,
\nonumber
\end{eqnarray}
where (for large $m$):
\begin{equation}\label{vcm}
v=\frac{(\vec{p}_n+2\vec{p}_{\bar{p}})^2}{4m}>0.
\end{equation}

Since the Faddeev equations will be solved for the S-wave, for comparison, we should also take the S-wave of the amplitude (\ref{an3d}) and of the amplitude
obtained from  (\ref{an3d})  by the replacement $u\to v$.
That is, we should calculate the integral $\frac{1}{2}\int_{-1}^1 \ldots dz$, where $z$ is the cosine of the angle between the vectors $\vec{p}_n$ and $\vec{p}_{\bar{p}}$.  In this way, depending on approximation, we find the S-wave amplitudes:\\
$
M(\bar{p}d\to e^+e^- n)=M_u\Gamma_0\\
$
or\\
$
M(\bar{p}d\to e^+e^- n)=M_v\Gamma_0,\\
$
where
\begin{eqnarray}
M_u&=&\frac{1}{2}\int_{-1}^1 dz \,\psi_d\left(\vert\vec{p}_n+\frac{1}{2}\vec{p}_{\bar{p}}\vert\right)\psi_{k_u}(0),
\label{an3u}
\\
M_v&=&\frac{1}{2}\int_{-1}^1 dz\, \psi_d\left(\vert\vec{p}_n+\frac{1}{2}\vec{p}_{\bar{p}}\vert\right)\psi_{k_v}(0).
\label{an3c}
\end{eqnarray}
$z$ enters both in 
$\vert\vec{p}_n+\frac{1}{2}\vec{p}_{\bar{p}}\vert$ and in $k_v$: 
\begin{eqnarray}
\vert\vec{p}_n+\frac{1}{2}\vec{p}_{\bar{p}}\vert&=&\sqrt{p_n^2+p_n p_{\bar{p}}z+\frac{1}{4}p^2_{\bar{p}}},
\label{arg}\\
k_v=\sqrt{m\,v}&=&\sqrt{\frac{1}{4}p_n^2+p_n p_{\bar{p}}z+p^2_{\bar{p}}},
\label{kv}
\end{eqnarray}
whereas $k_u$ does not depend on $z$:
\begin{equation}
k_u=\sqrt{m\,u}=\sqrt{\frac{3}{4}(\vec{p}\,^2_{\bar{p}}-\vec{p}\,^2_n)-m\vert\epsilon_d\vert}.
\label{ku}
\end{equation}
We would like to stress: according to \cite{DK81,DKK83}, the amplitude $M_u$ corresponds to the impulse approximation - Fig. \ref{fig1}a,
whereas the amplitude $M_v$ incorporates approximately all the graphs shown in Fig. \ref{fig1}.  

Note that $\psi_{k}(0)=1/f_{Jost}(-k)$, where $f_{Jost}(k)$ is the Jost function for the $\bar{p}p$ system. In numerical calculations we will use the Hulthen and exponential potentials, for which $\psi_{k}(0)$ is known analytically. It is given in Appendices \ref{Hulthen} and \ref{Exp}.

\section{Using Faddeev equations}\label{Fad}
For the two-body $p\bar{p}$ collision, the amplitude $\Gamma$ of annihilation $\bar{p}p\to e^+e^-$ is given by Eq. (\ref{Gamma_v}). It corresponds to the following transparent physics: the incident antiproton and the target proton, interacting with each other, are forming the two-body continuous 
spectrum state described by the wave function $\psi_k(r)$. This state itself does not include annihilation. Under annihilation interaction (taken in the first order),  the continuous spectrum state $\psi_k(r)$ turns into the state containing free $e^+e^-$ pair (neglecting their Coulomb interaction). Since the annihilation occurs at small distances, the transition amplitude is determined by $\psi_k(r=0)$ which appears in Eq. 
(\ref{Gamma_v}). 

The process  $\bar{p}d\to e^+e^-n$ is determined by the same physics: the  incident antiproton and the proton + neutron in the target deuteron,
interacting with each other, are forming the three-body continuous spectrum state described by the three-body wave function $\Psi(\vec{x},\vec{y})$ ($\vec{x},\vec{y}$ are the Jacobi coordinates). The annihilation turns the pair 
$\bar{p}p$ into  $e^+e^-$, i.e., the continuous spectrum state $\Psi(\vec{x},\vec{y})$ turns into the state containing three free particles $e^+e^-n$ (again neglecting the interaction between them). We will see that since the $\bar{p}p$ annihilation occurs at small distances, the transition amplitude is determined by $\Psi(\vec{x}=0,p_n)$, where $\vec{x}$ is the relative coordinate of the $\bar{p}p$ pair and the neutron momentum $p_n$ appears after Fourier transform in the second variable $\vec{y}$. 
The $p_n$-dependence in  $\Psi(\vec{x}=0,p_n)$ describes the momentum distribution of neutrons. 
This distribution is the only new principal element which differs the three-body problem from the two-body one. Though technically, the difference between the two-body and three-body problems is very considerable.

So, to calculate the amplitude of the reaction $\bar{p}d\to e^+e^-n$ we should first find the three-body continuous spectrum wave function 
$\Psi(\vec{x},\vec{y})$ of the $\bar{p}pn$ system  and then use it for calculating the amplitude of the transition into the  $e^+e^-n$ state. The three-body wave function is determined by the Faddeev equations with appropriate boundary (asymptotic) conditions containing the incident and outgoing deuteron + antiproton (in the case of elastic scattering only), or, in addition, outgoing baryonium $B$ ($\bar{p}p$ bound state) + neutron. For simplicity, we will consider the low incident antiproton momenta which are insufficient for the breakup -- to the $\bar{p}pn$ asymptotic state. That is, $p_{\bar{p}}<79.2$ MeV/c in the deuteron rest (lab.) frame, or $p_{\bar{p}}<52.8$ MeV/c in the $\bar{p}d$ c.m. frame. If the baryonium binding energy is greater than the deuteron one, the channel with the baryonium creation is open for any small momentum of the incident antiproton. 

Let us underline: the annihilation transforms one three-body state $\bar{p}pn$ (formed by the strong interaction) into another three-body one
$e^+e^-n$ (containing free particles). The $\bar{p}p$ annihilation can occur at any moment, not necessary in the asymptotic $\bar{p}pn$ state.  Therefore, to calculate this transition,  for the $\bar{p}pn$ system we will need just the full three-body continuous spectrum wave function $\Psi(\vec{x},\vec{y})$, not its asymptotic at $y\to\infty$. The asymptotic conditions will be imposed only to find this full wave function from the Faddeev equations. The asymptotic will be also used for tests. ({\it i})~We will check that the amplitudes \mbox{$\bar{p}d\to \bar{p}d$} and 
\mbox{$\bar{p}d\to Bn$} extracted from asymptotic satisfy  the unitarity condition involving both these amplitudes. This will confirm the full solution  $\Psi(\vec{x},\vec{y})$  found numerically. ({\it ii})~In some particular cases it turns out that the asymptotic starts rather early and reproduces with satisfactory precision the full wave function. With the analytical asymptotic form of the wave function one can find an analytical formula for the amplitude \mbox{$\bar{p}d\to e^+e^-n$}. We will check that in this case the amplitude found numerically is close to the analytical one. Besides, excluding from this formula the $\bar{p}p$ interaction (except for the $\bar{p}p$ annihilation), we reproduce the impulse approximation. This provides another test.
\bigskip

We will work in the coordinate space.  
The formulation of the Faddeev equations in the coordinate space see in Ref. \cite{RJ}.
The coordinate space three-body wave function is represented as sum of two Faddeev components:
\begin{eqnarray}\label{psi_i}
\Psi(\vec{x}_1,\vec{y}_1)&=&\psi_1(x_1,y_1)+\psi_3(x_3,y_3)
\\
&=&\frac{1}{\sqrt{4\pi}}\left(\frac{\chi_1(x_1,y_1)}{x_1 y_1}+\frac{\chi_3(x_3,y_3)}{x_3 y_3}\right).
\nonumber
\end{eqnarray}
Since the $\bar{p}n$ interaction is not taken into account (according to Refs.  \cite{DK81,DKK83}, it does not change the principal results), the Faddeev component $\psi_2(x_2,y_2)$  is absent. In each Faddeev component
we keep the S-wave only, therefore any Faddeev component depends on the modulas  of the corresponding Jacobi coordinates
defined in the Appendix \ref{Jacobi}. Whereas the modulas $x_3,y_3$ in  
$\psi_3(x_3,y_3)=\frac{1}{\sqrt{4\pi}}\frac{\chi_3(x_3,y_3)}{x_3y_3}$ in Eq. (\ref{psi_i}) are expressed via the vectors $\vec{x}_1,\vec{y}_1$ by Eqs. 
(\ref{xy31}) (see Appendix \ref{Jacobi}). This results in the angle integration in the r.h.-sides of the Faddeev equations (\ref{eq3f}) below.
We introduced in (\ref{psi_i}) the factor 
$\frac{1}{\sqrt{4\pi}}$ in order to have the conventional normalization (\ref{chi_norm}) for the radial bound state wave functions.

Since  $\Psi$ is finite, the functions 
$\chi_i(x,y)$ must satisfy the conditions
\begin{equation}\label{psi_0}
\chi_i(x=0,y)=\chi_i(x,y=0)=0. 
\end{equation}
Then the system of equations for the S-wave obtains the form:
\begin{eqnarray}\label{eq3f} 
&&\left[k^2+\frac{\partial^2}{\partial x^2} + 
\frac{\partial^2}{\partial y^2} - mV_{\bar{p}p}(x)\right]\chi_1(x,y)
\nonumber\\
&=&
\frac{1}{2}mV_{\bar{p}p}(x)\int_{-1}^{1}du \frac{xy}{x'y'}\chi_3(x',y'),
\nonumber\\
&&\left[k^2+\frac{\partial^2}{\partial x^2} + 
\frac{\partial^2}{\partial y^2} - mV_{pn}(x)\right]\chi_3(x,y)=
\nonumber\\
&=&\frac{1}{2}mV_{pn}(x)\int_{-1}^{1}du \frac{xy}{x'y'}\chi_1(x',y'),
\end{eqnarray} 
where $V_{\bar{p}p}(x),V_{pn}(x)$ are the $\bar{p}p$ and $pn$ potentials,
\begin{eqnarray}\label{xyp} 
x'&=&\frac{1}{2}(x^2+2\sqrt{3}xyu+3y^2)^{1/2},
\nonumber\\
y'&=&\frac{1}{2}(3x^2-2\sqrt{3}xyu+y^2)^{1/2},
\end{eqnarray}
and the Faddeev components $\chi_{1,3}$ have the asymptotic:
\begin{eqnarray}
\chi_1(x_1,y_1\to\infty)&=&\chi_b(x_1)\frac{1}{p_3}f_1(p_1)\exp(i p_1y_1),
\nonumber\\
&&
\label{eq61}\\
\chi_3(x_3,y_3\to\infty)&=&\chi_{d}(x_3)\frac{1}{p_3}
[\sin(p_3y_3)
\nonumber\\
&+&f_3(p_3)\exp(ip_3y_3)].
\label{eq63}
\end{eqnarray}
Here $\chi_b(x_1),\chi_d(x_3)$ are the wave functions of baryonium  ($\bar{p}p$) and deuteron both normalized as 
\begin{equation}\label{chi_norm}
\int_0^{\infty}\chi^2(r)dr=1.
\end{equation}
$f_1(p_1)$ is the dimensionless amplitude of the rearrangement $\bar{p}d\to Bn$.
$f_3(p_3)$ in (\ref{eq63}) is the dimensionless elastic $\bar{p}d$ scattering amplitude,  related to the phase shift as $f_3(p_3)=\frac{1}{2i}[\exp(2i\delta)-1]$. If the channel with rearrangement is closed, the phase $\delta$ is real.
$p_1$ is the Jacobi momentum in the $Bn$ channel, $p_3$ is the Jacobi momentum in the $\bar{p}d$ channel.
In these channels, in the c.m. frame,
 the Jacobi momenta are related to the c.m. ones as $p_{1}=\frac{\sqrt{3}}{2}\tilde{p}_n$, $p_3=\frac{\sqrt{3}}{2}p_{\bar{p}}$.
  In the elastic $\bar{p}d$ channel the momentum $p_{\bar{p}}$ is the incident (and final) antiproton momentum. In the rearrangement channel 
$\bar{p}d\to Bn$ the neutron momentum $\tilde{p}_n$ is found from the  energy conservation law. It is determined by the incident antiproton momentum $p_{\bar{p}}$ and by the particle masses participating in the reaction. Above we used the notation $\vec{p}_n$ (or $p_n$ without tilde) for the  neutron momentum in the reaction $\bar{p}d\to e^+e^-n$. It depends on the effective mass of the $e^+e^-$ pair and therefore varies. Whereas, 
$\tilde{p}_n$ is a particular value of $p_n$ which the neutron obtains in the reaction  $\bar{p}d\to Bn$, or kinematically equivalent, for the $e^+e^-$ effective mass equal to mass of baryonium. One should not confuse $\tilde{p}_n$ and $p_n$.

If the $\bar{p}p$ interaction is not enough  to create the  $\bar{p}p$ bound state, then 
\begin{eqnarray}
\chi_1(x_1,y_1\to\infty)&=&0,
\label{eq61_0}\\
\chi_3(x_3,y_3\to\infty)&=&\chi_{d}(x_3)\frac{1}{p_3}
[\sin(p_3y_3)
\nonumber\\
&+&f_3(p_3)\exp(ip_3y_3)],
\label{eq63_0}
\end{eqnarray}
and, as mentioned above, the phase shift corresponding to $f_3(p_3)$ is real.

The boundary conditions in the variables $x_1,x_3\to\infty$ correspond to decrease of both functions $\chi_1$ and $\chi_3$, due to finite size of the deuteron and baryonium.
That is:
\begin{equation}\label{x_large}
 \chi_1(x_1\to\infty,y_1)\to 0,\quad  \chi_3(x_3\to\infty,y_3)\to 0.
\end{equation}

Equivalently, instead  of Eqs. (\ref{x_large}), one can impose at $x_1,x_3\to\infty$ the equations (\ref{eq3f}) and put in them $V(x)=0$ (that we used in our numerical calculations). For negative energies the solutions of the free equations are zeroes, that is, are given by Eqs. (\ref{x_large}). 

If  the $\bar{p}p$ interaction is enough strong and creates the  $\bar{p}p$ bound state with rather large binding energy (but no excited states, for simplicity), the final relative $Bn$ momentum $p_1$ is rather large. The relative  $\bar{p}d$ momentum $p_3$  can be also large. Since  the oscillating factors 
$\exp(ip_1y_1)$ and $\exp(ip_3y_3)$, extracted in the asymptotic, are included in  $ \chi_1(x_1,y_1),\, \chi_3(x_3,y_3)$, the functions
(\ref{eq61}), (\ref{eq63})  oscillate. This complicates their numerical calculations.  A way to overcome that is to extract these exponents
from $ \chi_1(x_1,y_1),\, \chi_3(x_3,y_3)$. Besides, we have noticed that extraction of the deuteron and baryonium wave functions improves the accuracy
and convergence of the numerical calculations.   
That  is, we represent the Faddeev components as:
\begin{eqnarray}
 \chi_1(x_1,y_1)&=& \exp(ip_1y_1)\chi_b(x_1) \tilde{\chi}_1(x_1,y_1),
\label{chi1t} \\
 \chi_3(x_3,y_3)&=&\exp(ip_3y_3)\chi_d(x_3)\tilde{\chi}_3(x_3,y_3)
 \nonumber\\
 &+&\chi_d(x_3)\sin(p_3 y_3).
\label{chi3t}
\end{eqnarray}
We substitute expressions (\ref{chi1t}), (\ref{chi3t}) in the equations (\ref{eq61_0}), (\ref{eq63_0}), derive equations for 
the functions $\tilde{\chi}_1(x,y)$, $\tilde{\chi}_3(x,y)$, and impose on them the boundary conditions 
\begin{equation}
\tilde{\chi}_{1,3}(x,y=0)=0,\quad \left.\frac{\partial}{\partial y} \tilde{\chi}_{1,3}(x,y)\right\vert_{y=y_{max}}=0.
\label{cond1}
\end{equation}

Solving numerically the equations for  $\tilde{\chi}_{1,3}(x,y)$, we find the three-body wave function $\Psi(\vec{x},\vec{y})$, Eq. (\ref{psi_i}).

\section{Finding the  amplitude $\bar{p}d\to e^+e^-n$}\label{ampl}
As explained above,  the initial state containing  $\bar{p}d$, under strong interaction, turns into a continuous spectrum three-body
state $\bar{p}pn$ described by the wave function found in the previous section.  
Due to the $\bar{p}p$ annihilation, this state turns 
into the state of free particles $e^+e^-n$. To find the amplitude of this transition in the first order of the annihilation interaction, we should calculate the matrix element from the transition operator between the continuous spectrum state $\bar{p}pn$ and the final one $e^+e^-n$. This amplitude differs from the two-body one 
$\bar{p}p\to e^+e^-$ by the initial and final states, whereas the transition operator, corresponding to the annihilation, is the same. Since the two-body amplitude $\bar{p}p\to e^+e^-$ is known (it is given by Eq. (\ref{Gamma_u})), we can easy determine the  transition operator leading to this amplitude and then use this operator  in the three-body problem. Looking at the amplitude (\ref{Gamma_u}), we see that this operator has the form $\hat{O}=\delta^{(3)}(\vec{r})\Gamma_0$, where $\vec{r}$ is the relative distance between $\bar{p}$ and $p$ and $\Gamma_0$ is the $\bar{p}p\to e^+e^-$ annihilation amplitude not including the initial interaction via the $\bar{p}p$ potential. It corresponds to annihilation at $\vec{r}=0$. Since the pair $\bar{p}p$ after annihilation is absent in the final state and the $e^+e^-$ interaction in the final state is neglected,
the matrix element $\Gamma$, in the two-body reaction $\bar{p}p\to e^+e^-$, contains only the initial $\bar{p}p$ state, described by the continuous spectrum wave function $\psi_k(r)$:
\begin{equation}\label{G}
\Gamma=\Gamma_0\int \delta^{(3)}(\vec{r}) \psi_k(r) d^3r =\psi_k(r=0)\Gamma_0
\end{equation}
that indeed reproduces Eq. (\ref{Gamma_u}).

Similarly, for the three-body reaction $\bar{p}d\to e^+e^-n$, the initial state is given by the wave function $\exp(i\vec{P}\cd\vec{R})\Psi(\vec{x},\vec{y})$ with $\Psi$ defined in (\ref{psi_i}). The final state contains the (complex conjugated) neutron plane wave: 
$\exp(-i\vec{P}'\cd\vec{R})\psi_n^*$, where $\psi^*_n=\exp(-i \vec{p}_1\cd \vec{y}_1)$. The equations (\ref{eq61}), (\ref{eq63}), imply that the particle No. 1 is neutron, No. 2 -- proton and No. 3 -- antiproton. Therefore $\vec{y}_1$ is relative coordinate of neutron and $\bar{p}p$ system (apart the coefficient $\sqrt{3}/2$), $\vec{p}_1$ is the Jacobi momentum of the final neutron, that justifies the expression for $\psi_n$.
The factors $\exp(i\vec{P}\cd\vec{R})$ and $\exp(-i\vec{P}'\cd\vec{R})$ 
take into account the c.m. motion of the initial and final three-body systems. The center of mass coordinate $\vec{R}$ and the total momentum $\vec{P}$ are defined in Appendix \ref{Jacobi}, Eqs. (\ref{xyR3}) and (\ref{qpP3}) respectively.

In this way, for the amplitude of the transition $\bar{p}d\to e^+e^-n$ we find:
\begin{eqnarray}\label{M30}
&&M(\bar{p}d\to e^+e^-n)=\int \exp(-i\vec{P}'\cd\vec{R})\psi_n^*(y_1)
\nonumber\\
&\times&\hat{O}\exp(i\vec{P}\cd\vec{R})\Psi(\vec{x}_1,\vec{y}_1) d^3r_1d^3r_2 d^3r_3.
\end{eqnarray}
We transform the integration to the Jacobi variables:
$$
d^3r_1d^3r_2 d^3r_3 =\left(\frac{\sqrt{3}}{2}\right)^3 d^3x_1d^3y_1d^3R.
$$
After integration over $d^3R$ we get 
\begin{equation}\label{M_tot}
M(\bar{p}d\to e^+e^-n)=(2\pi)^3\delta^{(3)}(\vec{P}-\vec{P}')M_3\Gamma_0,
\end{equation}
where
\begin{eqnarray}\label{M3_0}
M_3&=&\left(\frac{\sqrt{3}}{2}\right)^3 \frac{1}{\sqrt{4\pi}}\int d^3x_1 d^3y_1e^{-i \vec{p}_1\cd\vec{y}_1} \delta^{(3)}(\vec{x}_1)
\nonumber\\
&\times&\left(\frac{\chi_1(x_1,y_1)}{x_1 y_1}
+\frac{\chi_3(x_3,y_3)}{x_3 y_3}\right).
\end{eqnarray}

We substitute here $x_3,y_3$ from (\ref{xy31}) and integrate over $\vec{x}_1$ by means of the delta-function $\delta^{(3)}(\vec{x}_1)$ that reduces the variables 
$\vec{x}_3,\vec{y}_3$ as follows:
\begin{eqnarray*}
&&\vec{x}_3=-\frac{1}{2}(\vec{x}_1+\sqrt{3}\vec{y}_1)\to -\frac{\sqrt{3}}{2}\vec{y}_1,
\\
&&\vec{y}_3=\frac{1}{2}(\sqrt{3}\vec{x}_1-\vec{y}_1)\to -\frac{1}{2}\vec{y}_1.
\end{eqnarray*}
We also integrate over the angles of the vector $\vec{y}_1$. In this way we obtain 
\begin{eqnarray}\label{M3}
M_3&=&\left(\frac{\sqrt{3}}{2}\right)^3 \frac{\sqrt{4\pi}}{p_1}\int_0^{\infty}dy_1\left[\lim_{x_1\to 0}\frac{1}{x_1}\chi_1(x_1,y_1)\right.
\nonumber\\
&+&\left.\frac{4}{\sqrt{3}y_1}\chi_3\left(\frac{\sqrt{3}}{2}y_1,\frac{1}{2}y_1\right)\right] \sin(p_1 y_1)
\nonumber\\
&=&\frac{\sqrt{3\pi}}{p_n} \int_0^{\infty}dy\left[\lim_{x\to 0}\frac{1}{x}\chi_1\left(x,\frac{2}{\sqrt{3}}y\right)\right.
\nonumber\\
&+&\left.\frac{2}{y}\chi_3\left(y,\frac{1}{\sqrt{3}}y\right)\right] \sin(p_n y).
\end{eqnarray}
Transforming the middle part of Eq. (\ref{M3}) into the r.h.-side, we made the replacement of the variable $y_1=\frac{2}{\sqrt{3}}y$ and took
 the values  $p_1=\frac{\sqrt{3}}{2}p_n$, $p_3=\frac{\sqrt{3}}{2}p_{\bar{p}}$ from Eq. (\ref{p1p3}) in Appendix \ref{Jacobi}. 

According to Eq. (\ref{eq61}), the function $\chi(x_1,y_1\to\infty)$  oscillates that complicates the numerical calculation of the first integral in the r.h.-side of (\ref{M3}). To overcome this difficulty, we represent $\chi(x_1,y_1)$ in the form: $\chi(x_1,y_1)=[\chi(x_1,y_1)-\chi^{as}(x_1,y_1)]+\chi^{as}(x_1,y_1)$, where $\chi^{as}(x_1,y_1)$ is the asymptotic form of $\chi(x_1,y_1)$ given by (\ref{eq61}). Then the integral containing the difference $[\chi(x_1,y_1)-\chi^{as}(x_1,y_1)]$ is calculated numerically without problem, whereas the integral containing $\chi^{as}(x_1,y_1)$ can be calculated analytically. Its analytical calculation is carried out in the next section. 

The amplitude $M_3$, Eq. (\ref{M3}), is the final result.  It determines the full amplitude $M(\bar{p}d\to e^+e^-n)$ by Eq. (\ref{M_tot}) and differs from it by a factor. $M_3$ corresponds to the case, considered in \cite{DK81,DKK83}, when both interactions -- $\bar{p}p$ and $pn$ -- are present (but the $\bar{p}n$ interaction is absent)  and 
is expressed via the three-body continuous spectrum wave function $\Psi(\vec{x},\vec{y})$ determined by the Faddeev equations. Since the Faddeev equations include all the  potential interactions, except for annihilation, expression (\ref{M3}) automatically takes into account, in addition to other re-scatterings, the initial $\bar{p}p$ interaction, which is described in (\ref{G}) by the factor $\psi_k(r=0)$.

\subsection{Amplitude $M_3$ for asymptotic wave function}\label{tr1}

In this subsection  we consider the case, when the Faddeev components in the three-body wave function $\Psi(\vec{x},\vec{y})$ are replaced by their asymptotic determined, in the S-wave, by the equations (\ref{eq61}), (\ref{eq63}). 
This allows to find analytically the pole in the amplitude related to creation of baryonium. In some cases, this can  also provide good numerical accuracy.

Substituting $\chi_1,\chi_3$ from (\ref{eq61}), (\ref{eq63}) into (\ref{M3}), we find $M_3$ in the following form:
$$
M_3=M_{31}+M_{33},
$$
where
\begin{eqnarray}\label{M31_0}
 M_{31}&=&\frac{\sqrt{3\pi}}{p_n }\int_0^{\infty}dy\lim_{x\to 0}\frac{1}{x}\chi_1(x,\frac{2}{\sqrt{3}}y)\sin(p_n y) 
\nonumber\\
&=&
\frac{\sqrt{4\pi}}{p_n p_{\bar{p}}}\lim_{x\to 0}\frac{1}{x}\chi_b(x)f_1\left(\frac{\sqrt{3}}{2}\tilde{p}_n\right) 
\nonumber\\
&\times&\int_0^{\infty}dy\exp(i\tilde{p}_ny)\sin(p_n y) 
\end{eqnarray}
To calculate the last integral, we regularize it, introducing in the integrand the factor $\exp(-\epsilon y)$ ($\epsilon>0$), and after integration take the limit $\epsilon\to 0$. That is,
\begin{eqnarray*}
&& \int_0^{\infty}dy\exp(-\epsilon y)\exp(i\tilde{p}_ny)\sin(p_n y)
 \\
  &=&
\frac{p_n}{{p}^2_n-(\epsilon -i\tilde{p}_n)^2}\vert_{\epsilon\to 0}
 =\frac{p_n}{{p}^2_n-\tilde{p}_n^2}.
\end{eqnarray*}
In this way, we find
\begin{equation}\label{M31a}
M_{31}=\frac{\sqrt{4\pi}}{p_n p_{\bar{p}}}\lim_{x\to 0}\frac{1}{x}\chi_b(x) f_1\left(\frac{\sqrt{3}}{2}\tilde{p}_n\right)\frac{p_n}{({p}^2_n-\tilde{p}^2_n)}.
\end{equation}
The contribution of the amplitude $M_{13}$ originates from the channel with the baryonium creation $\bar{p}d\to Bn$. We remind that 
$\tilde{p}_n$ is the neutron momentum in this channel. For given masses, it is fixed by the conservation laws. Whereas $p_n$ is the neutron momentum after annihilation, that is, in the reaction $\bar{p}d\to e^+e^-n$. Variation of the effective mass of the pair $e^+e^-$ corresponds to variation of $p_n$, whereas $\tilde{p}_n$ is the particular value of $p_n$ corresponding to the fixed effective $e^+e^-$ mass equal to mass of baryonium. 
According to Eq. (\ref{M31a}), $M_{31}$ vs. $p_n$ has the under-threshold pole corresponding to creation of baryonium.
 The baryonium creation amplitude $f_1$ is found from the Faddeev equations, which contain the Jacobi momenta.   Therefore $f_1=f_1(p_1)=f_1\left(\frac{\sqrt{3}}{2}\tilde{p}_n\right)$. We used the relation (\ref{p1p3}).
 
 Similarly, we find from (\ref{M3}) the contribution $M_{33}$:
\begin{eqnarray}\label{M33a}
M_{33}&=&\frac{2\sqrt{3\pi}}{p_n}\int_0^{\infty}\frac{dy}{y}\chi_3\left(y,\frac{1}{\sqrt{3}}y\right)\sin(p_ny)
\nonumber\\
&=&\frac{4\sqrt{\pi}}{p_n p_{\bar{p}}}\int_0^{\infty}\frac{dy}{y}
\chi_d(y) \left[\sin\left(\frac{1}{2}p_{\bar{p}} y\right)\right.
\nonumber\\
&+&\left.f_3\left(\frac{\sqrt{3}}{2}p_{\bar{p}}\right)\exp\left(\frac{i}{2}p_{\bar{p}} y\right)\right]\sin(p_n y).
\nonumber\\
&&
\end{eqnarray}
We again assumed that the amplitude $f_3$ is known as function of the Jacobi momentum, that is, 
$f_3=f_3(p_3)=f_3\left(\frac{\sqrt{3}}{2}p_{\bar{p}}\right)$.


\subsection{Without the $\bar{p}p$ interaction}\label{nopbarp}
In the present section we suppose that both the  $\bar{p}p$ and $\bar{p}n$ interactions are absent (though the annihilation 
$\bar{p}p\to e^+e^-$ is present). This exactly corresponds to the impulse approximation, Fig. \ref{fig1}a. The amplitude of reaction  $\bar{p}d\to e^+e^-n$  in the impulse approximation is obtained from Eqs. (\ref{an3u}) or (\ref{an3c}) by excluding from them the factor $\psi_k(0)$ responsible for the initial 
$\bar{p}p$ interaction. In this way, after omitting also the factor $\Gamma_0$, we find:
\begin{equation}\label{an3g}
\tilde{M}_S=\frac{1}{2}\int_{-1}^1 dz \psi_d\left(\vert\vec{p}_n+\frac{1}{2}\vec{p}_{\bar{p}}\vert\right). 
\end{equation}
The amplitude obtained above by the Faddeev formalism, after omitting the $\bar{p}p$ interaction and $\Gamma_0$, should coincide with (\ref{an3g}). 
This will be a test of our calculation.

In absence of the $\bar{p}p$ interaction we have $f=f_1=0$. Therefore,   from (\ref{M31a}) we get $M_{31}=0$ and from (\ref{M33a}) we obtain:%
\begin{equation}\label{M3a}
M_{33}=\frac{2\sqrt{\pi}}{p_n p_{\bar{p}}}\int_0^{\infty}\frac{dy}{y}
\chi_d(y)\sin\left(\frac{1}{2}p_{\bar{p}}y\right) \sin\left(p_n y\right).
\end{equation}

To prove the coincidence of the amplitude (\ref{an3g}) and (\ref{M3a}),
at first, we express $\psi_d(p)$ in (\ref{an3g}) via the coordinate space wave function
\begin{eqnarray*}
\psi_d(p)&=&\int \psi_d(r)\exp(i\vec{p}\vec{r})d^3r
\\
&=&\sqrt{4\pi}\int_0^{\infty} dr\chi_d(r) \frac{\sin(pr)}{p}.
\end{eqnarray*}
Then Eq. (\ref{an3g}) obtains the form:
\begin{equation}\label{an3e}
\tilde{M}_S=
\sqrt{4\pi}\int_0^{\infty} dr\chi_d(r)\frac{1}{2}\int_{-1}^1 dz \frac{\sin(\vert\vec{p}_n+\frac{1}{2}\vec{p}_{\bar{p}}\vert r)}
{\vert\vec{p}_n+\frac{1}{2}\vec{p}_{\bar{p}}\vert}.
\end{equation}
With expression  (\ref{arg}) for $\vert\vec{p}_n+\frac{1}{2}\vec{p}_{\bar{p}}\vert$, the integral (\ref{an3e}) over $z$ is calculated explicitly. 
The result exactly
coincides with (\ref{M3a}). Since Eq. (\ref{M3a}) was obtained as a particular case from the general expression (\ref{M3}), 
the latter one is also confirmed in this way.

\section{Numerical results and discussion}\label{num}
The derivation of the result found in Refs. \cite{DK81,DKK83} -- our equation (\ref{an3c}) -- is general, it does not rely on any particular model of $np$ and $\bar{p}p$ interaction.
Therefore, it should be valid for any $np$ and $\bar{p}p$ potentials. The numerical calculations, presented in this section, were carried out with three potentials: the Hulthen potential, the exponential one and the spherically symmetric potential well. We chose them since they provide the analytical solutions for the two-body  bound and scattering state wave functions. We can take the unrealistic potentials, since we don't intend to compare with experimental data, but we will compare the amplitudes found from the Faddeev equations with Eqs. (\ref{Mu}) and (\ref{Mv}). Since we will find the coincidence with Eq. (\ref{Mv}), we check this coincidence with three different potentials, to show that this coincidence is not by chance.
That's quite enough for our aim. Therefore, the fact that the potentials are unrealistic does not cast a shadow on our conclusions.
The form of  potentials was taken the same for the $np$ and $\bar{p}p$ systems, but the parameters are different. Depending on the system, we chose  the parameters of the potential to provide the correct deuteron binding energy ($-2.23$ MeV) and the realistic
$\bar{p}p$ binding energy (two cases: $-15$ and $-55$ MeV).  As mentioned, for all these potentials, there are explicit solutions of the two-body problem - the bound state wave functions, used for solving the Faddeev equations,
and the scattering state wave functions, determining $\psi_k(r=0)$ in Eqs. (\ref{an3u}), (\ref{an3c}). 
The results are similar for all three potentials. Therefore, we will mainly present the results found with the Hulthen potential, though    
we will also give some results found with
the exponential potential. The Hulthen potential and corresponding wave functions are presented in Appendix \ref{Hulthen}.  
The exponential potential and the corresponding wave functions are given in Appendix \ref{Exp}.

As explained above, to calculate the amplitude of the reaction $\bar{p}d\to e^+e^-n$, solving the Faddeev equations, we need full three-body wave function 
$\Psi(\vec{x},\vec{y})$, not only its asymptotic. To check this solution, we take the asymptotic tails of this wave function, extract from them the elastic
($\bar{p}d\to \bar{p}d$) and rearrangement ($\bar{p}d\to Bn$) scattering amplitudes and check, whether they satisfy the unitarity conditions. The details of this procedure and the numerical results are presented in Appendix \ref{unitarity}. We consider two cases: ({\it i}) when the $\bar{p}p$  interaction is not enough to create baryonium (the elastic scattering only) and ({\it ii}) when the $\bar{p}p$  interaction   creates baryonium (the elastic scattering and rearrangement). 

In the case ({\it i}), the test of unitarity is reduced to the test that the phase shift $\delta$, extracted from the scattering amplitude, is real. The results for two antiproton incident momenta $p_{\bar{p}}=23.9$ MeV/c and $p_{\bar{p}}=48.3$ MeV/c are given in the Table \ref{tab1} in Appendix \ref{unitarity}. They show that the phase shift is indeed real with high accuracy. 

In the case ({\it ii}) (two open channels), the test of unitarity is reduced to the test of the equality (\ref{eqf}). The results for the same two antiproton incident momenta are given in the Table \ref{tab2} in Appendix \ref{unitarity}. They show that in this case, in our numerical calculation,  the unitarity is satisfied with accuracy better than 0.1\%. 

These tests confirm that our numerical solutions of the Faddeev equations are correct.

\begin{figure}[h!]
\begin{center}
\includegraphics[width=7.5cm]{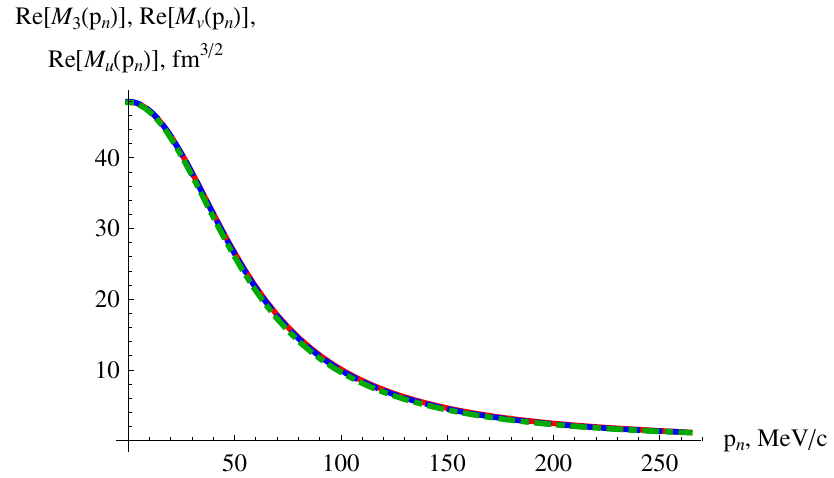}
\\
\includegraphics[width=7.5cm]{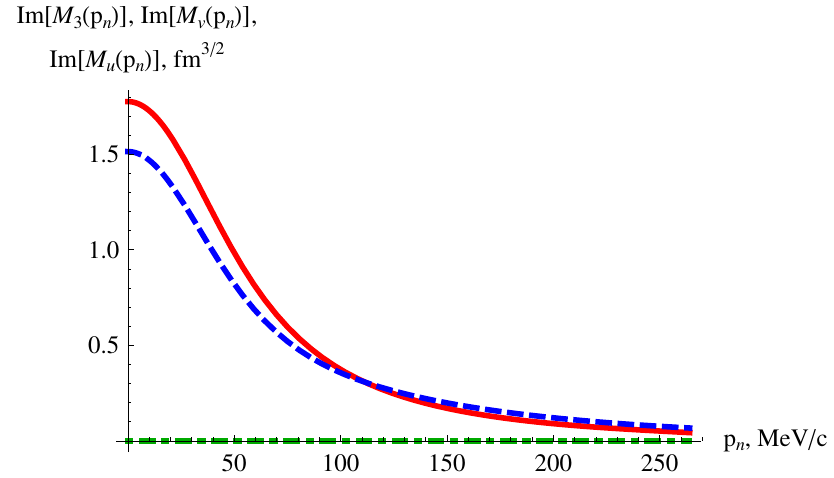}
\end{center}
\caption{(Color online) Amplitude of the reaction $\bar{p}d\to e^+e^-n$, divided by the annihilation amplitude $\Gamma_0$ for the incident c.m. antiproton momentum $p_{\bar{p}}=48.3$ MeV/c vs. outgoing c.m. neutron momentum $p_n$. No $\bar{p}p$ bound state. 
\\
Upper panel: Real part of the amplitude. Solid curve --  calculated via Faddeev formalism, Eq. (\ref{M3}). Dashed curve  -- calculated by Eq. (\ref{an3c}) with $k=\sqrt{m\,v}$ given by Eq. (\ref{kv}). Dot-dashed curve -- impulse approximation, calculated by Eq. (\ref{an3u}) with 
$k=\sqrt{m\,u}$ given by Eq. (\ref{ku}). The curves are indistinguishable from each other.
\\
Lower panel: the same as the upper one but for imaginary part of amplitude.}
\label{fig4}
\end{figure}

\begin{figure}[h!]
\begin{center}
\includegraphics[width=7.5cm]{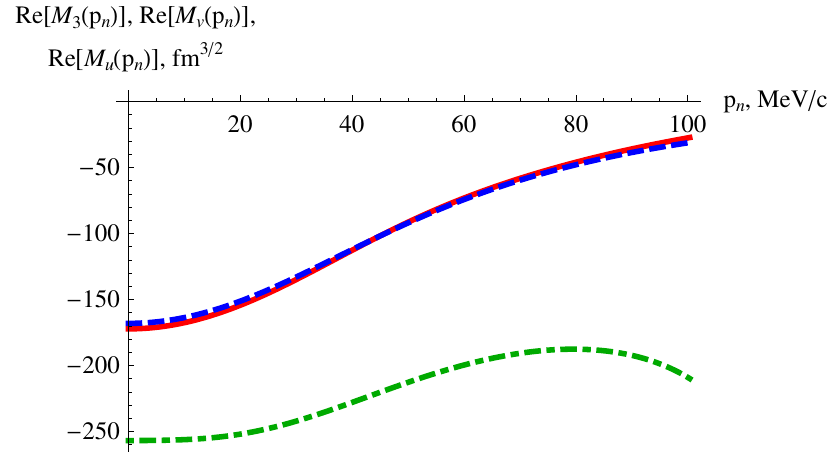}
\includegraphics[width=7.5cm]{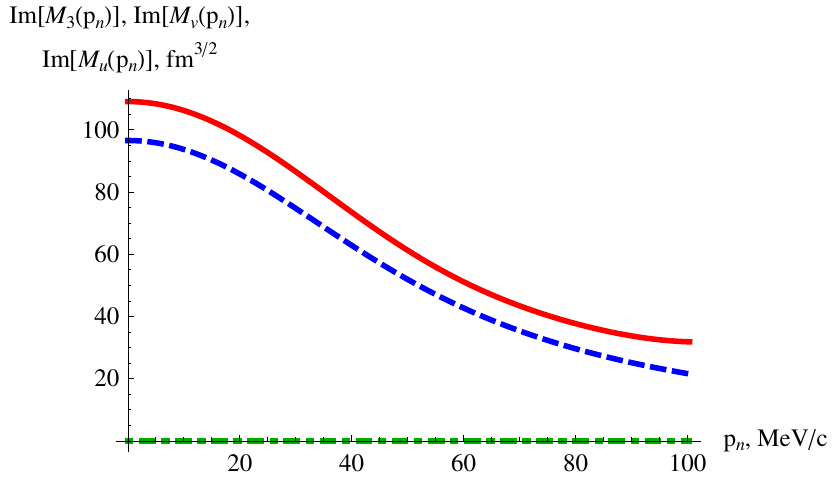}
\end{center}
\caption{The same as in Figures \ref{fig4} for $p_{\bar{p}}=48.3$ MeV/c. The $\bar{p}p$ pair is bound with $\vert \epsilon_b\vert =15$ MeV.
\label{fig6}}
\end{figure}

\begin{figure}[h!]
\begin{center}
\includegraphics[width=7.5cm]{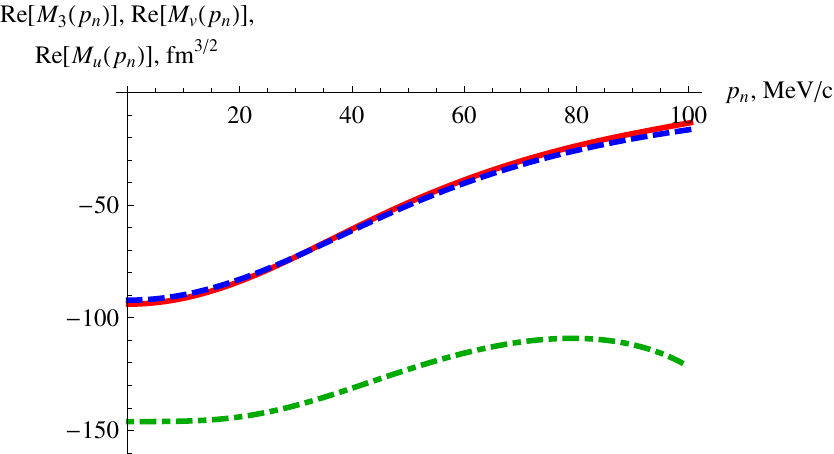}
\includegraphics[width=7.5cm]{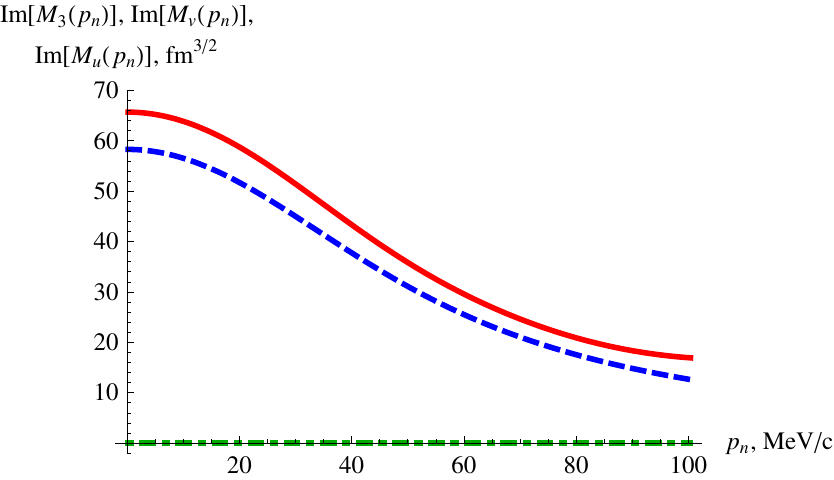}
\end{center}
\caption{The same as in Figures \ref{fig6} but for the exponential potential given in Appendix \ref{Exp}.
\label{fig6a}}
\end{figure}

\begin{figure}[h!]
\begin{center}
\includegraphics[width=7.5cm]{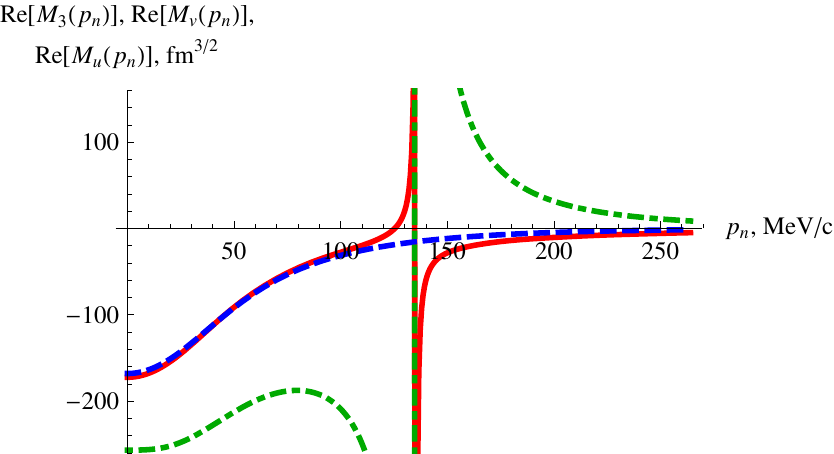}
\includegraphics[width=7.5cm]{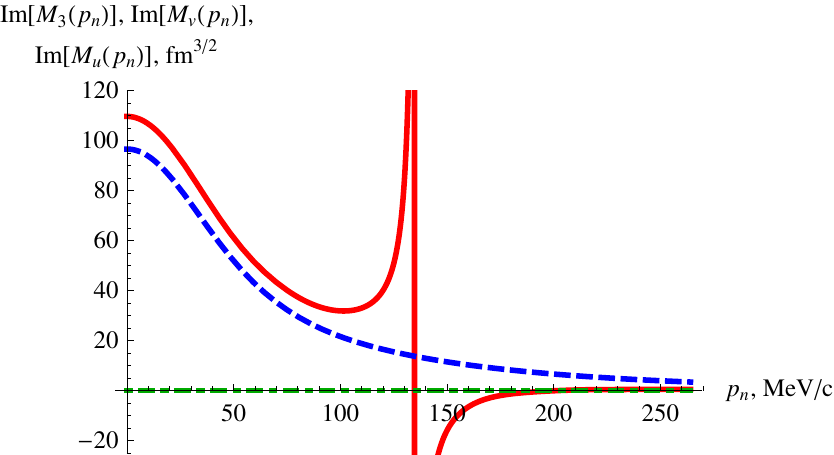}
\end{center}
\caption{The same as in Figures \ref{fig6} for $p_{\bar{p}}=48.3$ MeV/c, but for larger domain of $p_n$, covering the response from the $\bar{p}p$ bound 
state with $\vert \epsilon_b\vert =15$ MeV. 
\label{fig62}}
\end{figure}

\begin{figure}[h!]
\begin{center}
\includegraphics[width=7.5cm]{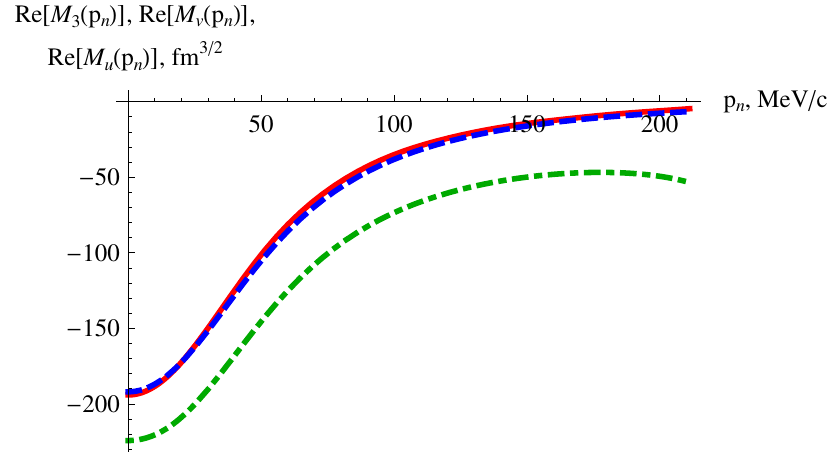}
\includegraphics[width=7.5cm]{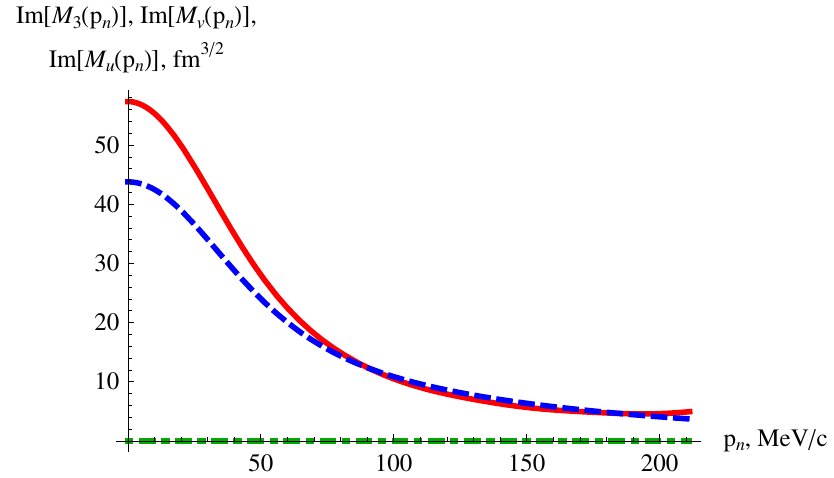}
\end{center}
\caption{The same as in Figures \ref{fig4} and \ref{fig6} for $p_{\bar{p}}=48.3$ MeV/c. The $\bar{p}p$ pair is bound with $\vert \epsilon_b\vert =55$ MeV.
\label{fig8}}
\end{figure}

\begin{figure}[h!]
\begin{center}
\includegraphics[width=7.5cm]{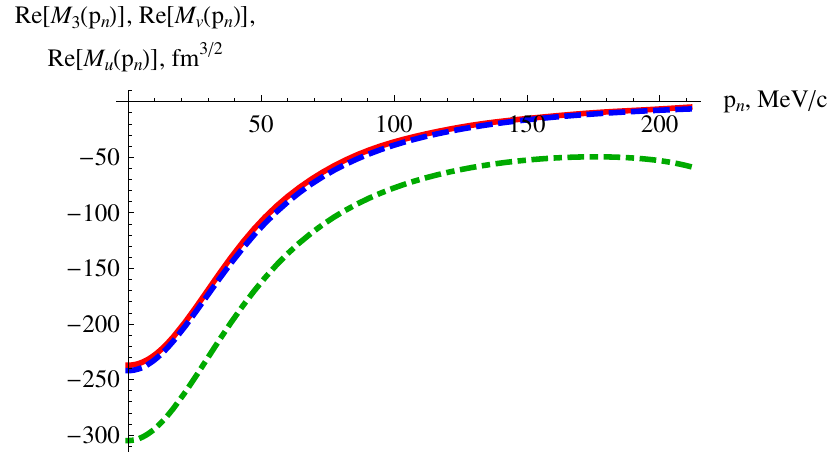}
\\
\includegraphics[width=7.5cm]{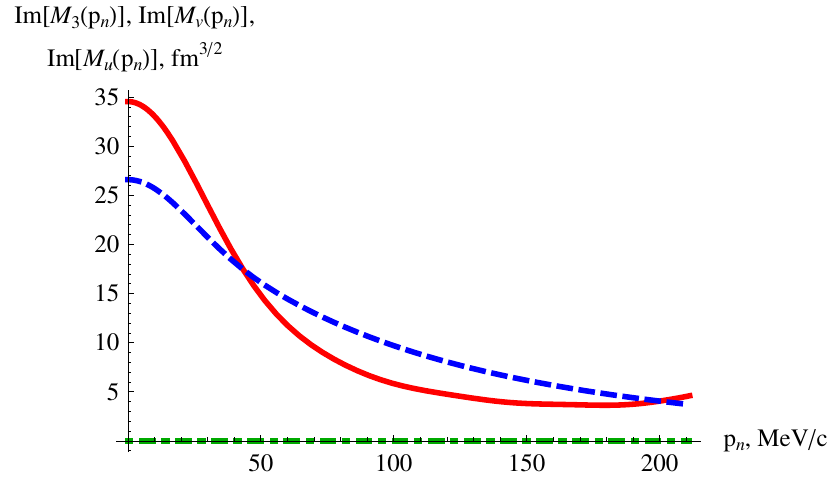}
\end{center}
\bigskip
\caption{The same as in Figures \ref{fig4} - \ref{fig6} and \ref{fig8}, but for $p_{\bar{p}}=23.9$ MeV/c. The $\bar{p}p$ pair is bound with $\vert \epsilon_b\vert =55$ MeV.}
\label{fig7}
\end{figure}

In Figures \ref{fig4}---\ref{fig7}, the real and imaginary parts of the amplitude of reaction $\bar{p}d\to e^+e^-n$, however, not including the factor 
$\Gamma_{0}$, for the incident antiproton momentum $p_{\bar{p}}=48.3$ MeV/c (Figs. \ref{fig4}-\ref{fig8})  and $p_{\bar{p}}=23.9$ MeV/c (Fig. \ref{fig7}) are shown as function of the outgoing neutron momentum $p_n$. The calculations for Figs. \ref{fig4}--\ref{fig6} and  Figs. \ref{fig62}--\ref{fig7} were fulfilled with the Hulthen potential (Appendix \ref{Hulthen}), whereas the curves shown in Fig. \ref{fig6a} were calculated with the exponential potential (Appendix \ref{Exp}).
 Everywhere the solid curve is the Faddeev calculation by Eq. (\ref{M3}), the dashed curve is calculated by the prescription of Refs. \cite{DK81,DKK83}, in which we keep the S-wave only,
i.e., by Eq. (\ref{an3c}), the dot-dashed curve is the impulse approximation Eq. (\ref{an3u}).

 The curves in Fig. \ref{fig4} were calculated for the potential parameter $V_{0\bar{p}p}=10$ MeV. 
This interaction is not enough  to create the  $\bar{p}p$ bound state.  
In the three-body $\bar{p}pn$ problem, the elastic scattering channel $\bar{p}d\to\bar{p}d$ only is open. 

We see that in Figure \ref{fig4} (upper panel) all three curves - Faddeev calculation, Eq. (\ref{M3}), the one calculated by Eq. (\ref{an3c}) (with $k_v=\sqrt{m\,v}$, given by Eq. (\ref{kv})) and impulse approximation -- calculated by Eq. (\ref{an3u}) (with $k_u=\sqrt{m\,u}$, given by Eq. (\ref{ku})), are practically indistinguishable from each other. This is, apparently, a consequence of weak $\bar{p}p$ interaction allowing to neglect 
re-scattering of $\bar{p}$ on the deuteron nucleons which just makes the difference between the three cases. Hence, the diagram Fig. \ref{fig1}a dominates in this case. 

In contrast to the reals parts, the imaginary parts of the amplitudes -- lower panel of Fig. \ref{fig4}, calculated by the Faddeev equations and by Eq. (\ref{an3c}), visibly differ from each other. However, the imaginary parts in Fig. \ref{fig4} are at least 20 times smaller than the real ones. 
Whereas, the accuracy of  Eq. (\ref{an3d}) was estimated in  \cite{DK81,DKK83} in 30\%. In impulse approximations, the imaginary parts are zero.
So, the magnitudes of the imaginary parts are beyond of this accuracy and, therefore, cannot be reproduced, whereas the reals parts are described by Eq. (\ref{an3c}) even with better accuracy than it was expected in  \cite{DK81,DKK83}. It can be also noticed that the larger the imaginary parts, the better they are reproduced through the Faddeev equations.

Note that replacing in the full three-body wave function by its asymptotic, i.e., calculating the sum of amplitudes by Eqs. (\ref{M31_0}) 
and (\ref{M33a}) (the case considered in Sec. \ref{tr1}), we obtain the results, both for real and imaginary parts (not shown), which are close to the solid curves in Fig. \ref{fig4}. This means that for these parameters asymptotic of the three-body wave function starts rather early.

The main result of the present work is illustrated by Figs. \ref{fig6}-\ref{fig7}. In these figures, the case when the $\bar{p}p$ interaction is sufficient for the formation of baryonium is considered.  Figure \ref{fig6} corresponds to the case when a $\bar{p}p$ bound state in the potential --- baryonium --- with energy  $\vert\epsilon_b\vert=15$ MeV, provided by  the potential parameters given in Appendix \ref{db_H}. In Fig. \protect{\ref{fig62}} the domain of the neutron momentum $p_n$ is shown in which the $\bar{p}p$ bound state clearly manifests itself.
 Figures \ref{fig8} and \ref{fig7} correspond to the baryonium binding energy  
$\vert\epsilon_b\vert=55$ MeV. 
We remind that everywhere  
$r_0=0.6061$ fm. The curves in Figures \ref{fig4}-\ref{fig8} are calculated for the incident antiproton momentum $p_{\bar{p}}=48.3$ MeV/c.  Figure \ref{fig7} corresponds to  $p_{\bar{p}}=23.9$ MeV/c.


For different $\bar{p}p$ binding energies and incident momenta of antiproton, for the Hulthen potential,  the real parts of the amplitudes calculated by the approximate formula derived in \cite{DK81,DKK83}, i.e., by Eq. (\ref{an3c}), with $k_v$ given by Eq. (\ref{kv}), are well reproduced by the "exact" calculations via Faddeev  equations. 
There is some deviation for the imaginary parts, which, however, is within the accuracy of Eq. (\ref{an3d}), estimated in \cite{DK81,DKK83}. 

One can wonder
if the agreement shown in our work is not accidental in the sense that it may not append for another parameters of the Hulthen potential or another functional form. To check that, we carried out the calculations with another potential (the exponential one) given in Appendix \protect{\ref{Exp}} and providing quite different wave function.
As it is seen in Fig. \ref{fig6a}, the values of amplitudes calculated with the exponential potential differs from ones found with the Hulthen potential, but the relative positions of the curves is practically the same. This shows that our conclusions remain unchanged under considerable variation of the $pn$ and $\bar{p}p$ interaction.

The same situation is reproduced for the spherically symmetric potential well (not shown). Though in the latter case, the deviation of imaginary parts is a little bit larger. 
Note that the imaginary parts of the amplitudes calculated in the impulse approximation Eq. (\ref{an3u}) (the dot-dashed curves at the right panels in all the figures) is zero.

The coincidence of the curves found, following Refs. \cite{DK81,DKK83}, by  Eq. (\ref{an3c}), with the results obtained by the Faddeev equations, 
is  even better than one can expect from estimation of the precision made in  \cite{DK81,DKK83}.  These two coinciding results - solid and dashed curves - considerably differ from the impulse approximation  - the dot-dashed curve. This shows that the series of graphs Fig. \ref{fig1} is indeed important. Whereas, the delicate cancellations between the non-adiabatic and off-shell effects, really take place. Due to these cancellations, the sum of this series is reduced to the impulse approximation amplitude with the shifted argument $k_u\to k_v$ except for vicinity of the $\bar{p}p$ bound state, i.e., in the interval $p_n=130\div 140$ MeV/c in Fig. \ref{fig62}. In this vicinity the condition (1) from  \cite{DK81}: $\frac{1}{3}\vert \Gamma_u-\Gamma_v\vert \gg \vert \Gamma_v\vert$ is satisfied that, as indicated in Refs. \cite{DK81,DKK83}, violates the applicability of Eq. (\ref{an3c}).

\section{Conclusion}\label{concl}
The annihilation amplitude $\bar{p}p\to e^+e^-$ of free $\bar{p}p$ pair is graphically shown in Fig. \ref{fig2}. According to Eq. (\ref{Gamma_u}), it is given by the product $\psi_k(0)\Gamma_0$, where the factor $\psi_k(0)$ incorporates the $\bar{p}p$ interaction in the initial state, whereas 
$\Gamma_0$ is the "pure" annihilation amplitude which does not take into account the initial state interaction between $\bar{p}$ and $p$. The factor 
 $\psi_k(0)$ provides important information about the $\bar{p}p$ interaction. It was hope that from the reaction  $\bar{p}d\to e^+e^-n$, where the effective
 $e^+e^-$ mass can take the under-threshold values (less than $2m$), one can extract 
 the time-like proton form factor in the under-threshold domain. As it follows from the results of Refs. \cite{DK81,DKK83}, confirmed by the present work, this hope is not realized in full measure.
 As Figure \ref{fig62} shows, the existence and position of the $\bar{p}p$
 bound states indeed can be found.  However, the replacement in the initial state of proton by deuteron results in the set of much more complicated  graphs, shown in Fig. \ref{fig1}, which, in general, are not reduced to the factor $\psi_k(0)\Gamma_0$. As it was shown in Refs. 
\cite{DK81,DKK83} analytically,  due to cancellations of the non-adiabatic and off-mass-shell effects in these graphs, 
the sum of graphs Fig. \ref{fig1} is reduced to the product $\psi_{k_v}(0)\Gamma_0$, where $\Gamma_0$ still depends on the momentum transfer squared $q^2$, which can take the under-threshold values $q^2<4m^2$, but everywhere, except for  the close vicinity of the bound state, the factor $\psi_{k_v}(0)$ is determined by the energy (\ref{vcm}) which is always positive. 
That is,  the positions of the $\bar{p}p$  bound states can be found in the reaction $\bar{p}d\to e^+e^-n$. Whereas, 
the values of the proton electromagnetic form factor under threshold remain inaccessible, since the reaction $\bar{p}d\to e^+e^-n$ provides the factor 
$\psi_{k_v}(0)$, determining the proton time-like form factor for the $\bar{p}p$ energy $v$ which is not under but above threshold.

To check this result and the cancellation of the  non-adiabatic and off-mass-shell effects,
we have calculated numerically the amplitude of the reaction $\bar{p}d\to e^+e^-n$, treating the three-body system $\bar{p}pn$ by means of the Faddeev equations.  Everywhere, except for vicinity of singularities corresponding to the $\bar{p}p$  bound states,
the result of this calculation turned out to be rather close to the amplitude calculated by Eq. (\ref{an3c}), derived in Refs. \cite{DK81,DKK83}. In this way, we confirm the conclusion of Refs. \cite{DK81,DKK83}. Simultaneously, we confirm the delicate cancellations of the  non-adiabatic and off-mass-shell effects in the graphs shown in Fig. \ref{fig1}.

We emphasize again that in Eqs. (\ref{an3d}), (\ref{an3u}), the $\bar{p}p$ bound states manifest themselves in the under-threshold behavior of the wave function $\psi_{k_u}(r=0)=\frac{1}{f_{Jost}(-k_u)}$, where $f_{Jost}(k)$ is the Jost function. However, everywhere except for close vicinity of the $\bar{p}p$ bound state
the re-scatterings and cancellation of the non-adiabatic and off-shell effects change the value of $\psi_{k_u}(0)$, resulting in the shift $k_u\to k_v$, where the energy $v$ is always positive that makes the proton under-threshold form factor inaccessible in this domain in the reaction $\bar{p}d\to e^+e^-n$.
Closeness  of the solid and dashed curves in the figures shown in Sec. \ref{num} confirms numerically the cancellation of the 
non-adiabatic and off-shell effects proved analytically in  \cite{DK81,DKK83}.

In Refs. \cite{DK81,DKK83} an experimental manifestation of the expression (\ref{Mv}) for the $\bar{p}d\to e^+e^-n$ amplitude was discussed (Fig. 3 from \cite{DK81} and Fig. 7 from \cite{DKK83}). It requires the experiments in which the neutron is emitted in two opposite directions relative to momentum of incident antiproton. 
Solving Faddeev equations, we restricted our calculations by the S-waves only that corresponds to the replacement of Eq. (\ref{Mv}) by (\ref{an3c}).
Therefore, in our approximation, the amplitude is insensitive to the  emitted neutron momentum direction.
However, in our simplified calculations, nothing particular related to the S-wave was used to test eq. (\ref{an3c}). We expect it should be valid for the higher partial waves too. In our opinion, the calculations presented in this paper and confirming Eq. (\ref{Mv}) for the S-wave increase the interest to the experimental test proposed in \cite{DK81,DKK83} which involve the higher partial waves.

What would  happen if we could gradually weaken the $\bar{p}p$ interaction so that the $\bar{p}p$ re-scatterings 
become unimportant? For example, if $V_0\to 0$ in the Hulthen potential, Eq. (\ref{Hulthen_pot}). We would come in the  situation considered in Sec. \ref{nopbarp}. In this case, the impulse approximation dominates without any shift $k_u\to k_v$. This does not prevent from possibility to access the under-threshold proton form factor. However, the weak $\bar{p}p$ interaction does not create any bound states or resonances and does not change significantly the "bare" form factor $\Gamma_0(q^2)$. It would be still possible to go down under threshold, but, in this case, there is no $\bar{p}p$ interaction there. The Jost function (\ref{jost0_h}) (where $B\to 0$) and the factor $\psi_k(0)$ tend to 1. So, the same reason - enough strong $\bar{p}p$ interaction, which creates the  baryonium, prevents from access to the proton form factor (except for access to $\Gamma_0(q^2)$) outside the vicinities of the $\bar{p}p$ bound states  in the reaction $\bar{p}d\to e^+e^-n$. If this reason disappears, access becomes possible, but it will be access to the desert from point of view of the $\bar{p}p$ interaction.
\bigskip

\bmhead{Acknowledgments} The authors are grateful to \mbox{J.~Carbonell} and \mbox{R.~Lazauskas} for useful recommendations on solving the Faddeev equations.

\appendix
\section{Hulthen potential}\label{Hulthen}
For convenience, we will assume that both the proton and neutron and proton and antiproton interact by means of the Hulthen potential
\begin{equation}\label{Hulthen_pot}
 V(r)=-V_0\frac{\exp(-\frac{r}{r_0})}{1-\exp(-\frac{r}{r_0})}=-\alpha\mu\frac{\exp(-\mu r)}{1-\exp(-\mu r)},
\end{equation}
($V_0=\alpha\mu$, $r_0 =1/\mu\leftrightarrow \alpha=V_0r_0$, $\mu=1/r_0$), with the same parameter $r_0$, however, with different parameters $V_0$.
We assume that all these particles have equal masses $m$. 
The  wave function  for the S-wave in the coordinate space  is represented as 
$$
\psi_k(r)=\frac{1}{\sqrt{4\pi}}\frac{\chi_k(r)}{r}. 
$$
with
\begin{equation}\label{chi}
\chi(r)=N e^{-\kappa r}(1-e^{-\mu r}),
\end{equation}
where $\kappa=\frac{1}{2}(\alpha m-\mu)=\sqrt{B_1m}$ and
$
N=\sqrt{2\kappa (\kappa + \mu) (2 \kappa + \mu)}/ \mu.
$
This wave function is normalized by Eq. (\ref{chi_norm}).

The ground state binding energy for two particles of the equal masses $m$ reads  (see e.g. \cite{flugge}, Problem 68:
\begin{equation}\label{B1}
B_1=\vert E_1\vert=\frac{1}{4m}(\alpha m-\mu)^2.
\end{equation}
The bound state exists if $\alpha m>\mu$. Everywhere $m$ {\it is not} reduced mass, but mass of one of the particles.

For the Hulthen potential
 the Jost function  $f_{Jost}(k)$ is known in analytical form \cite{Newton}:
 \begin{eqnarray}\label{jost0_h}
f_{Jost}(k)&=&
\frac{\Gamma\left(1 + 2iA\right)}
{\Gamma\left(1+iA - \sqrt{B^2-A^2}\right)}
\nonumber\\
&\times&
\frac{1}{\Gamma\left(1+iA + \sqrt{B^2-A^2}\right)}.
\end{eqnarray}
where we denoted:
$$
A=\sqrt{mE} r_0=k r_0=\frac{k}{\mu},\; B^2=mV_0 r^2_0=\alpha \frac{m}{\mu}.
$$
$\Gamma(z)$ is the gamma-function.

As well known:
\begin{equation}\label{an3b}
\psi_{k}(r=0)=\frac{1}{f_{Jost}(-k)}.
\end{equation}

\subsection{Deuteron and baryonium parameters}\label{db_H}
For the parameters $\hbar=197.3$ MeV$\cdot$fm, $m=938.3$ MeV, $r_0=0.6061$ fm, $\epsilon_d=-2.23$ MeV (deuteron),  we find 
$V_0=144.7$  MeV.

For the same parameters except for $\epsilon_b=-15$ MeV (baryonium),  we find $V_0=195.3$  MeV.

For the same parameters except for $\epsilon_b=-55$ MeV (baryonium),  we find $V_0=270.6$  MeV.

\section{The exponential potential}\label{Exp}
In the exponential potential
\begin{equation}\label{VE1}
V(r)=-V_0\exp\left(-\frac{r}{r_0}\right)
\end{equation}
like in the Hulthen one, there are explicit analytical solutions for 
the ground and scattering state wave functions, as well as the Jost function \cite{Newton},  Ch. 14.

The bound state wave function reads:
\begin{equation}\label{VE3}
\chi(r)=AJ_{\tilde{\nu}}\left[\alpha\exp\left(-\frac{r}{2r_0}\right)\right],
\end{equation}
where $\tilde{\nu}=2r_0\sqrt{m\vert E\vert},\;\alpha=2r_0\sqrt{mV_0}$ and
  $J_{\tilde{\nu}}$ is the Bessel function of the first kind.

The eigenvalue equation follows from $\chi(0)=0$. It obtains the form
\begin{equation}\label{VE2}
J_{2r_0\sqrt{m\vert E\vert}}\left(2r_0\sqrt{mV_0}\right)=0
\end{equation}

The Jost function $f(k)$ reads\footnote{The Jost function (\ref{VE7})
differs from Eq. (14.9) in chapter 14, ref. \cite{Newton}, by the complex conjugation.} \cite{GW}:
\begin{eqnarray}\label{VE7}
f_{Jost}(k)&=&\left(\frac{\alpha}{2}\right)^{-\nu}J_{\nu}(\alpha)\Gamma(\nu+1)
\nonumber\\
&=&\exp[-ir_0k\log(r_0^2mV_0)]
\nonumber\\
 &\times&J_{2ir_0k}\left(2r_0\sqrt{mV_0}\right)\Gamma(2ir_0k+1).
 \nonumber\\
\end{eqnarray}

The value $\psi_{k}(r=0)$ is given by Eq. (\ref{an3b}). Therefore:
\begin{equation}\label{VE8}
\psi_{k}(r=0)=\frac{\exp\left[-ir_0k\log\left(r_0^2mV_0\right)\right] }{J_{-2ir_0k}\left(2r_0\sqrt{mV_0}\right)\Gamma(-2ir_0k+1)}.
\end{equation}

\subsection{Deuteron and baryonium wave functions}\label{db_E}
For the parameters $\hbar=197.3$ MeV$\cd$fm, $m=938.3$ MeV, $r_0=0.6061$ fm, $\epsilon_d=-2.23$ MeV (deuteron),  we find 
$V_0=225.6$  MeV.
The deuteron wave function obtains the form ($r$ in $fm$, $\chi(r)$ in $fm^{-1/2}$):
\begin{equation}\label{eq9E}
\chi_d(r)=0.7023J_{\nu}(y),
\end{equation} 
where $\nu=0.2810,\; y=2.826\exp(-0.8249r)$.

For the same parameters except for $\epsilon_b=-15$ MeV (baryonium),  we find $V_0=338.4$  MeV.
The   baryonium wave function reads ($r$ in $fm$, $\chi(r)$ in $fm^{-1/2}$):
\begin{equation}\label{eq10E}
\chi_b(r)=1.175J_{\nu}(y),
\end{equation} 
where $\nu=0.7288,\; y=3.462\exp(-0.8249r)$.

Both wave functions are normalized to 1 by Eq. (\ref{chi_norm}).

\section{Three-body Jacobi coordinates and momenta}\label{Jacobi}
Three-body relative coordinates are defined as: 
\begin{eqnarray}\label{xyR3}
&&\vec{x}_3=\vec{r}_1-\vec{r}_2,\quad
\vec{y}_3=\frac{2}{\sqrt{3}}\left(\frac{\vec{r}_1+\vec{r}_2}{2}-\vec{r}_3\right),
\nonumber\\
&&\vec{R}=\frac{\vec{r}_1+\vec{r}_2+\vec{r}_3}{3}.
\end{eqnarray}
Two other sets of relative coordinates are obtained  by cyclic permutation. In particular:
\begin{eqnarray}\label{xyR1}
&&\vec{x}_1=\vec{r}_2-\vec{r}_3,\quad
\vec{y}_1=\frac{2}{\sqrt{3}}\left(\frac{\vec{r}_2+\vec{r}_3}{2}-\vec{r}_1\right),
\nonumber\\
&&\vec{R}=\frac{\vec{r}_1+\vec{r}_2+\vec{r}_3}{3}.
\end{eqnarray}
The set $\vec{x}_3,\vec{y}_3$ is expressed via  $\vec{x}_1,\vec{y}_1$:
\begin{equation}\label{xy31}
\vec{x}_3=-\frac{1}{2}(\vec{x}_1+\sqrt{3}\vec{y}_1), \quad \vec{y}_3=\frac{1}{2}(\sqrt{3}\vec{x}_1-\vec{y}_1).
\end{equation}
Inverse relations:
\begin{eqnarray}\label{inv}
&&\vec{r}_1=\vec{R}+\frac{1}{2}\vec{x}_3+\frac{1}{2\sqrt{3}}\vec{y}_3,
\nonumber\\
&&\vec{r}_2=\vec{R}-\frac{1}{2}\vec{x}_3+\frac{1}{2\sqrt{3}}\vec{y}_3,\quad
\vec{r}_3=\vec{R}-\frac{1}{\sqrt{3}}\vec{y}_3.
\nonumber\\
\end{eqnarray}

Substituting Eqs. (\ref{inv}) in the scalar product, we find:
$$
\vec{k}_1\cd\vec{r}_1+\vec{k}_2\cd\vec{r}_2+\vec{k}_3\cd\vec{r}_3= 
\vec{q}_3\cd\vec{x}_3+\vec{p}_3\cd\vec{y}_3+\vec{P}\cd\vec{R},
$$
where 
\begin{eqnarray}\label{qpP3}
&&\vec{q}_3=\frac{1}{2}(\vec{k}_1-\vec{k}_2),\quad \vec{p}_3=\frac{1}{\sqrt{3}}\left(\frac{\vec{k}_1+\vec{k}_2}{2}-\vec{k}_3\right),
\nonumber\\
&&\vec{P}=\vec{k}_1+\vec{k}_2+\vec{k}_3.
\end{eqnarray}
and similarly for another set:
\begin{eqnarray}\label{qpP1}
&&\vec{q}_1=\frac{1}{2}(\vec{k}_2-\vec{k}_3),\quad \vec{p}_1=\frac{1}{\sqrt{3}}\left(\frac{\vec{k}_2+\vec{k}_3}{2}-\vec{k}_1\right),
\nonumber\\
&&\vec{P}=\vec{k}_1+\vec{k}_2+\vec{k}_3.
\end{eqnarray}
which provides:
$$
\vec{k}_1\cd\vec{r}_1+\vec{k}_2\cd\vec{r}_2+\vec{k}_3\cd\vec{r}_3= 
\vec{q}_1\cd\vec{x}_1+\vec{p}_1\cd\vec{y}_1+\vec{P}\cd\vec{R},
$$
We will use only these coordinates and momenta.
We don't use any other combinations called the "relative"\, coordinates and momenta. However, we will also use the particle momenta themselves defined in the c.m. frame of reaction.

Let us express the Jacobi momenta $\vec{p}_1$, $\vec{p}_3$ through the c.m. ones. Let the  particle 1 is the neutron and the particle 2 is the proton and the particle 3 is the antiproton. That is, in the c.m. frame: $\vec{k}_1=\vec{p}_n, \;\vec{k}_2=\vec{p}_p$ and 
$\vec{k}_3=\vec{p}_{\bar{p}}$. In these notations:  
\begin{eqnarray}\label{p13}
\vec{p}_1&=&\frac{1}{\sqrt{3}}\left(\frac{\vec{p}_p+\vec{p}_{\bar{p}}}{2}-\vec{p}_n\right),
\nonumber\\
\vec{p}_3&=&\frac{1}{\sqrt{3}}\left(\frac{\vec{p}_n+\vec{p}_p}{2}-\vec{p}_{\bar{p}}\right).
\end{eqnarray}
In the c.m. frame: $\vec{p}_n+\vec{p}_p+\vec{p}_{\bar{p}}=0\to \vec{p}_p=-\vec{p}_n-\vec{p}_{\bar{p}}$. Therefore:
\begin{equation}\label{p1p3}
\vec{p}_1=-\frac{\sqrt{3}}{2}\vec{p}_n,\quad \vec{p}_3=-\frac{\sqrt{3}}{2}\vec{p}_{\bar{p}}.
\end{equation}

\section{Test of unitarity}\label{unitarity}
The fulfillment of unitarity  provides very strong test of the solution of the Faddeev equations.
We take the incident antiproton momentum which is not enough for the deuteron breakup. We will consider two cases: ({\it i}) the $\bar{p}p$ system has no bound states; ({\it ii}) the $\bar{p}p$ system has bound state (baryonium). 

In the case ({\it i}), the $\bar{p}d$ scattering is elastic. Hence, the corresponding phase shift $\delta$ is real. In this case, the test of unitarity is reduced to the test that $\delta$ is real. The amplitude $f_3(p_3)$ in Eq. (\ref{eq63_0}) reads 
$f_3(p_3)=\frac{1}{2i}(\exp(2i\delta)-1)$ (we use the definition of $f$ without momentum $p_3$ in the denominator). 
Then $\exp(2i\delta)=1+2if_3(p_3)$. We will check that $\vert \exp(2i\delta)\vert =1$, that is
\begin{equation}\label{eqf0}
\vert 1+2if_3(p_3)\vert=1.
\end{equation}

The results for  $p_{\bar{p}}=23.9,\; 48.3$ MeV/c  are given in the Table \ref{tab1}.
\begin{table}[h]
\begin{center}
\begin{minipage}{174pt}
\caption{The amplitude $f_3(p_{\bar{p}})$ for elastic $\bar{p}d$ scattering for $V_{0\bar{p}p}=10$ MeV and $r_0=0.6061$ fm}\label{tab1}%
\begin{tabular}{@{}lll@{}}
\toprule
$p_{\bar{p}}$, MeV/c  & 23.9 & 48.3\\
\midrule
$f_3(p_{\bar{p}})$&  $0.0202+0.0004i$& $0.0370+0.0014i$\\
$2if_3(p_{\bar{p}})+1$&  $0.9992+0.0404i$& $0.9972+0.0740i$\\
$\vert 2if_3(p_{\bar{p}})+1\vert$&  $1.0000$& $0.9999$\\
\botrule
\end{tabular}
\end{minipage}
\end{center}
\end{table}

The last line of this table contains the values  $\vert \exp(2i\delta)\vert =\vert 1+2if_3(p_3)\vert$. 
We see that $\vert\exp(2i\delta)\vert\approx 1$ with high accuracy. This proves that $\delta$ is real, that confirms the validity of the numerical   solution of the Faddeev equation.

In the case of open channel with the baryonium creation the value $\vert \exp(2i\delta)\vert$ is smaller than 1.  This indicates that the part of the final particles goes in this channel. Inclusion of this channel should restore the unitarity condition. This channel is taken into account below.

We deal now with the 6D space (two the 3D Jacobi coordinates $\vec{X}=(\vec{x}_i,\vec{y}_i)$). For $(\vec{x}_i,\vec{y}_i)$ one can chose any pair of the Jacobi coordinates.  Unitarity means the conservation of flax:
\begin{equation}\label{j3}
\vec{j}=\frac{i\hbar}{2m}(\Psi\vec{\nabla}_{x_i}\Psi^*-\Psi^*\vec{\nabla}_{x_i}\Psi+\Psi\vec{\nabla}_{y_i}\Psi^*-\Psi^*\vec{\nabla}_{y_i}\Psi),
\end{equation}
where $\vec{\nabla}_{x_i}=\frac{\partial}{\partial \vec{x}_i}$, $\vec{\nabla}_{y_i}=\frac{\partial}{\partial \vec{y}_i}$.
The conservation of flax is expressed by the formula
\begin{equation}\label{int}
\oint \vec{j} d\vec{S}=0,
\end{equation}
where $d\vec{S}$ is element of surface in the 6D space. We integrate over the sphere of large radius $\rho=\sqrt{\vec{x}^2+\vec{y}^2}\to\infty$, where the three-body wave function is known and is given by its asymptotic form. In this way, Eq. (\ref{int}) provides a relation (the unitarity condition) between the amplitudes $f_3,f_1$. It has the form:
\begin{equation}\label{eqf}
\vert 1+2if_3(p_3)\vert^2=1-\frac{4p_1}{p_3}\vert f_1(p_1)\vert^2.
\end{equation}
If the rearrangement channel is closed, that is, $f_1=0$, the equation (\ref{eqf}) turns into (\ref{eqf0}).
\begin{table}[!h]
\begin{center}
\begin{minipage}{174pt}
\caption{The amplitude $f_3(p_{3})$ and $f_1(p_{1})$ for $\vert \epsilon_b\vert=55$ MeV, $V_{0\bar{p}p}=270.6$ MeV and $r_0=0.6061$ fm}\label{tab2}%
\begin{tabular}{@{}lll@{}}
\toprule
$p_{\bar{p}}$, MeV/c  & 23.9 & 48.3\\
\midrule
$p_3$, MeV/c &  $20.7$& $41.8$\\
$p_1$, MeV/c &  $223.5$& $226.3$\\
$f_3(p_{3})$&  $-0.13+0.02i$& $-0.25+0.07i$\\
$f_1(p_{1})$&  $-0.00-0.01i$& $-0.02-0.02i$\\
$\vert 1+ 2if_3(p_{3})\vert^2$&  $0.9886$& $0.9763$\\
$1-\frac{4p_1}{p_3}\vert f_1(p_1)\vert^2$&  $0.9879$& $0.9761$\\
\botrule
\end{tabular}
\end{minipage}
\end{center}
\end{table}

In the last two lines of the Table \ref{tab2} we see that the values $\vert 1+2if_3(p_3)\vert^2$ and $1-\frac{4p_1}{p_3}\vert f_1(p_1)\vert^2$ coincide with each other within the precision better than 0.1\%.


\end{document}